\documentclass[twocolumn]{pasj00}                                                       
\SetRunningHead{Okumura~et~al.}{SN~Ia rates with SXDS}

\usepackage{times}

\begin{document}

\title{The Type~Ia supernovae rate with Subaru/XMM-Newton Deep Survey}
\author{
Jun E. Okumura\altaffilmark{1,3},
Yutaka Ihara\altaffilmark{2,3}, 
Mamoru Doi\altaffilmark{2,4}, 
Tomoki Morokuma\altaffilmark{2}, 
Reynald Pain\altaffilmark{5},
Tomonori Totani\altaffilmark{3,4,1},
Kyle Barbary\altaffilmark{6,7},
Naohiro Takanashi\altaffilmark{8},
Naoki Yasuda\altaffilmark{9},
Greg Aldering\altaffilmark{7},
Kyle Dawson\altaffilmark{7,10},
Gerson Goldhaber\altaffilmark{7, \dag},
Isobel Hook\altaffilmark{11,12},
Chris Lidman\altaffilmark{13},
Saul Perlmutter\altaffilmark{6,7}, 
Anthony Spadafora\altaffilmark{7},
Nao Suzuki\altaffilmark{7, 9}, and
Lifan Wang\altaffilmark{14}
}
\altaffiltext{1}{Department of Astronomy, School of Science, Kyoto University, Sakyo-ku, Kyoto 606-8502}
\altaffiltext{2}{Institute of Astronomy, University of Tokyo, 2-21-1 Osawa, Mitaka-shi, Tokyo 181-0015}
\altaffiltext{3}{Department of Astronomy, Graduate School of Science, University of Tokyo, 7-3-1 Hongo, Bunkyo-ku, Tokyo 113-0033}
\altaffiltext{4}{Research Center for the Early Universe, Graduate School of Science, University of Tokyo, 7-3-1 Hongo, Bunkyo-ku, Tokyo 113-0033}
\altaffiltext{5}{LPNHE, CNRS/IN2P3, Universit{\'e} Pierre et Marie Curie, Universit{\'e} Denis Diderot, 4 place Jussieu, 75252 Paris Cedex 05, France}
\altaffiltext{6}{Department of Physics, University of California Berkeley, Berkeley, CA 94720, USA}
\altaffiltext{7}{Lawrence Berkeley National Laboratory, 1 Cyclotron Road, Berkeley, CA 94720, USA}
\altaffiltext{8}{The University of Tokyo Executive Management Program, Ito International Research Center 2nd floor, 7-3-1 Hongo, Bunkyo-ku, Tokyo 113-0033}
\altaffiltext{9}{Kavli Institute for the Physics and Mathematics of the Universe (IPMU), The University of Tokyo, Kashiwano-ha 5-1-5, Kashiwa-shi, Chiba 277-8568}
\altaffiltext{10}{Department of Physics and Astronomy, University of Utah, Salt Lake City, UT 84112, USA}
\altaffiltext{11}{University of Oxford Astrophysics, Denys Wilkinson Building, Keble Road, Oxford OX1 3RH, UK}
\altaffiltext{12}{Osservatorio Astronomico di Roma, via Frascati 33, 00040 Monteporzio (RM), Italy}
\altaffiltext{13}{Australian Astronomical Observatory, PO Box 915, North Ryde NSW 1670, Australia}
\altaffiltext{14}{Physics Department, Texas A\&M University, College Station, TX 77843, USA}
\email{okumura@kusastro.kyoto-u.ac.jp}

\maketitle

\begin{abstract}
  We present measurements of the rates of high-redshift Type~Ia
  supernovae derived from the Subaru/XMM-Newton Deep Survey (SXDS).
  We carried out repeat deep imaging observations with Suprime-Cam on
  the Subaru Telescope, and detected 1040 variable objects over 0.918~deg$^2$ 
  in the Subaru/XMM-Newton Deep Field.  From the imaging
  observations, light curves in the observed $i'$-band are
  constructed for all objects, and we fit the observed light curves
  with template light curves.  Out of
  the 1040 variable objects detected by the SXDS,
  39 objects over the redshift range $0.2 < z < 1.4$
  are classified as Type~Ia supernovae using the light curves.
  These are among the most distant SN~Ia rate measurements to date. 
  We find that the Type~Ia supernova rate increase up to $z \sim 0.8$ and may then flatten at higher redshift.
  The rates can be fitted by a simple power law, $r_V(z)=r_0(1+z)^\alpha$ with
  $r_0=0.20^{+0.52}_{-0.16}$(stat.)$^{+0.26}_{-0.07}$(syst.)$\times 
   10^{-4} {\rm yr}^{-1}{\rm Mpc}^{-3}$, and
   $\alpha=2.04^{+1.84}_{-1.96}$(stat.)$^{+2.11}_{-0.86}$(syst.).
\end{abstract}

\footnotetext[\dag]{Deceased July 19 2010}

\section{Introduction}

Type~Ia supernovae (SNe~Ia) are remarkable objects as cosmological
distance indicators, having provided the first direct evidence of the
accelerating cosmic expansion. This cosmic acceleration was first
reported by two independent supernova observation teams: the Supernova
Cosmology Project (SCP) (Perlmutter~et~al. 1999) and the High-Z SN
Search Team (Riess~et~al. 1998).  Since then, many large SN surveys
have been carried out to accurately measure the cosmological
parameters (e.g., Knop~et~al. 2003; Tonry et~al. 2003; Astier~et~al. 2006; Riess et 
al. 2007; Wood-Vasey~et~al. 2007; Kowalski~et~al. 2008; Hicken et
al. 2009; Amanullah~et~al. 2010; Sullivan~et~al. 2011; Suzuki~et~al. 2012). 

Although SNe~Ia are effective as standard candles, their
 progenitors are yet to be conclusively identified. It is widely 
believed that the progenitor of a SN~Ia is a binary system
containing a C+O white dwarf, and recently the compact nature of the 
exploding star has been confirmed (Nugent~et~al. 2011; Bloom~et~al. 2012).
There are two widely discussed scenarios for the progenitor, 
the single degenerate (SD) scenario and the double degenerate (DD) scenario. 
In the SD scenario, a C+O white dwarf accretes gas from a
companion star in a binary system. Its mass increases up to the
Chandrasekhar limit where it explodes as an SN~Ia (e.g., Nomoto 1982;
Hachisu~et~al. 1996; Nomoto~et~al. 1997).  If SNe~Ia from the SD
scenario exist, the companion star survives in the supernova remnant
after the SN~Ia explosion. Various methods have been used to search
for such companion stars, but to date no clear detection has been made
(e.g., Ruiz-Lapuente~et~al. 2004; Ihara~et~al. 2007;
Gonzalez~et~al. 2009; Kerzendorf~et~al. 2009; Schaefer \& Pagnotta
2012; Li~et~al. 2011; Nugent~et~al. 2011; Bloom~et~al. 2012;
Brown~et~al. 2012; Chomiuk~et~al. 2012; Margutti~et~al.
2012).  Other constraints on the SD scenario had been obtained from 
radio observations, which showed no clear evidence of an interaction 
between the ejecta and the circumstellar material (CSM) surrounding the SNe.
 (e.g. Panagia~et~al. 2006; Hancock~et~al. 2011; Chomiuk~et~al. 2012).
Though these observations disfavor non-degenerate donors, some SN~Ia
spectra show narrow time varying and/or blue-shifted Na I D absorption features 
possibly associated with a SD donor star (e.g. Patat~et~al. 2007; Simon~et~al. 2009; Blondin~et~al. 2009; Stritzinger~et~al. 2010; Sternberg~et~al. 2011; Maguire~et~al. 2013). 
These features are also investigated in the context of DD scenario (Shen~et~al. 2013; Raskin \& Kasen 2013).

In the DD scenario, a merger of two C+O white dwarfs with a combined
mass exceeding the Chandrasekhar mass leads to an SN~Ia explosion
(e.g., Iben \& Tutukov 1984; Webbink 1984).  Searches have been
carried out to detect DD binaries that could be SN~Ia progenitors,
but strong limits have not yet been reached due to small number statistics 
(e.g., Koester~et~al. 2005; Geier~et~al. 2007).

SNe~Ia explode with a ``delay time'' between binary system formation
and subsequent SN explosion. This delay time is one of the primary
methods for understanding the progenitor scenario of SNe~Ia.  Recent
comparisons of the SN~Ia rate and the cosmic star formation history
(e.g., Madau~et~al. 1998; Hopkins \& Beacon 2006) have resulted in a
wide range of derived delay times (e.g., Cappellaro~et~al. 1999; Pain
et~al. 2002; Tonry~et~al. 2003; Barris \& Tonry 2006;
Neill~et~al. 2006; Poznanski~et~al. 2007; Botticella~et~al. 2008; 
Dahlen~et~al. 2008; Dilday~et~al. 2008; Kuznetsova~et~al. 2008
Graur~et~al. 2011, Maoz~et~al. 2012, Barbary~et~al. 2012,
Perrett~et~al. 2012).  The delay time can be predicted theoretically
from the SD scenario or the DD scenario (see Wang \& Han 2012, for a
review).  For example, in the SD scenario, the delay time is closely
related to the lifetime of the companion star (e.g. Hachisu~et~al. 2008). 
The delay time obtained from rate studies can allow us to distinguish 
between the two progenitor scenarios.

Many studies have derived the SN~Ia delay time distribution (DTD) from
observations.  Totani~et~al. (2008), for example, measured the
delay time distribution based on the stellar age estimate of each
galaxy in a sample of passively evolving SXDS galaxies, finding
that the time distribution could be described by a featureless power law 
going as $f_D (t) \propto t^{-1.08\pm0.15}$ over $t$ = 0.1$-$10 Gyr. 
Other studies using different methods also show consistency with a 
$t^{-1}$ trend (see Maoz \& Mannucci 2012, for a review).

In this context, high-redshift SN~Ia rates ($z>1$) play an important 
role in investigating the DTD, especially for the short delay time regime. 
If SNe~Ia with short delay times dominate SN~Ia populations, 
the cosmic SN~Ia rate evolution should closely trace that of
the cosmic star formation, and thus high-redshift SN~Ia rates provide
information about the short delay time population.

High-redshift SN~Ia rates have been measured in several surveys.
Recently, Perrett~et~al. (2012) has measured the SN~Ia rate over the
redshift range $0.1\le z \le 1.1$ using 286 spectroscopically
confirmed and $\gtrsim$ 400 photometrically identified SNe~Ia from the
Supernova Legacy Survey (SNLS).  Dahlen~et~al. (2008; hereafter Da08)
obtained the first SN~Ia rate measurement beyond $z$ of 1, based on
the Great Observatories Origins Deep Survey (GOODS) survey. They used
56 high-redshift SNe~Ia, the majority of which were spectroscopically
confirmed.  Interestingly, they find that the SN~Ia rate decreases
beyond $z \sim 1.6$, contradicting the expectation from the delay time
distribution measurements of Totani~et~al. 2008. However, the rate in
the highest redshift bin ($z \gtrsim 1.4$) has a large uncertainty due
to small number statistics. The detection efficiency at $z>1.4$
rapidly decreases with redshift as the observed bands shift farther
into the rest-frame UV. Other high-z rate measurements 
have been reported in the literature. 
Graur~et~al. 2011 derived the SN~Ia rate up to $z \sim 2.0$ using 150 SNe 
from a SN survey in the Subaru Deep Field (SDF), and found that the
SN~Ia rate levels off at $1.0 < z< 2.0$. Their SN
classification method is based on a single epoch
in the $R, i', $and $z'$ bands, provided in Poznanski~et~al. 2007.
Barbary~et~al. 2012 (hereafter B12) derived the SN~Ia rate up to $z \sim 1.6$ using
$\sim 20$ SNe~Ia from the {\it Hubble Space Telescope} Cluster
Supernova Survey, finding a rate that is broadly consistent with previous 
measurements but with large uncertainties. The behavior of SN~Ia
rates at high redshifts is not clear yet due to the large statistical
uncertainties associated with the few detections to date in this
redshift range. This is a key issue in SN~Ia rate studies.

In this paper, we measure the SN~Ia rate to high redshift using the
Subaru/XMM-Newton Deep Survey (SXDS) data set.  The survey area is
large ($\sim$1~deg$^2$) enough to obtain many SNe~Ia.  With the
repeat imaging observations we are able to construct high-quality SN
light curves.  We obtain 39 SNe~Ia in the range $0.2 \lesssim z
\lesssim 1.4$ using a classification method that relies primarily on
light-curve fitting and photometric redshifts. Spectroscopic identifications, 
source colors, host galaxy redshifts, and X-ray data are employed when 
available to directly remove contamination or to improve statistical contamination corrections.

\begin{figure}
\centering
 \FigureFile(80mm,60mm){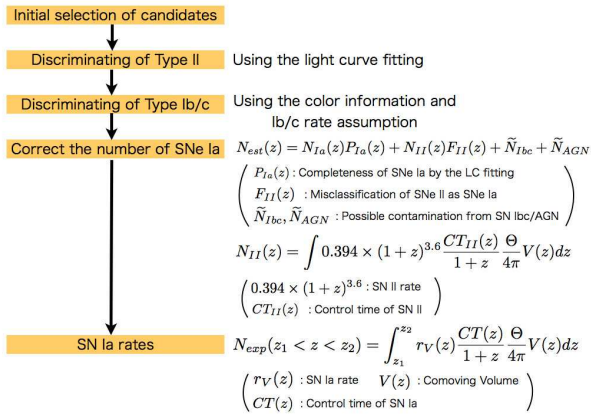}
\caption{The flow chart of the rate calculation. Since the SN
  classification has a bias, we need to correct the number of SNe~Ia
  identified by their light curves using artificial light curves
  made by the Monte Carlo simulations.
}\label{flow}
\end{figure}

The paper is organized as follows. In \S\ref{observations}, we describe the 
observations. We briefly summarize the SXDS data set, and describe in
detail the observations used for SN detection. In \S\ref{sn}, we describe
the SN selection method. In \S\ref{ratecalc}, we describe the control time calculation. 
In \S\ref{results}, we discuss the results and estimate systematic errors.
We provide Figure~\ref{flow} as a visual guide to the procedure
used in this paper. In \S\ref{discussion}, we compare our 
results with previous works.
Finally, we summarize our work in  \S\ref{conclusions}. Throughout
the paper we adopt the cosmological parameters $H_0 = 70$ km s$^{-1}$
Mpc$^{-1}$, $\Omega_M=0.3$, $\Omega_\Lambda=0.7$. All magnitudes are
given in the AB system.\\

\section{Observations} \label{observations}
\subsection{Imaging observations} \label{imaging}

The Subaru/XMM-Newton Deep Survey (SXDS) is a multi-wavelength survey
from X-ray to radio (Sekiguchi~et~al. 2004)
The survey targets a 1.22 deg$^2$ field centered on ($02^{\rm
  h}18^{\rm m}00^{\rm s}$,$-05^{\circ}00'00''$), hereafter referred to
as the Subaru/XMM-Newton Deep Field (SXDF). The optical imaging
component of the survey was carried out using Suprime-Cam (Miyazaki~et~al. 2002) 
on the 8.2--m Subaru telescope, starting in September 2002.
With Suprime-Cam's very wide field of view ($34' \times 27'$), the
field is covered in five pointings (SXDF-C, SXDF-N, SXDF-S, SXDF-E and
SXDF-W; see Furusawa~et~al. 2008).  In order to detect and follow the
light curves of optically faint variable objects, the Suprime-Cam
observations were split into exposures of 1800$-$7200 seconds
separated by periods of days to weeks.  Between September 2002 and
December 2002, the fields were observed 5$-$7 times in the $i'$-band
and 2$-$4 times in the $R_c$- and $z'$-bands.  After the 2002
observations finished, we took reference images in the $i'$-band in
2003 and 2005, in the $z'$-band in 2005, and in the $R_c$-band in
2008. The observations are described in detail in Morokuma~et~al. (2008a, hereafter M08). 
Here, the number of the epochs of the observations, each
exposure time, and each detection limit are summarized in Table~1. For
our SN study we exclude regions around bright objects to ensure
reliable detection of object variability. This reduces the total
effective area to 0.918~deg$^2$.

X-ray imaging observations in the SXDF were carried
out with the European Photon Imaging Camera (EPIC) on the
XMM-Newton. The X-ray imaging covers most of the SXDF fields. 
The X-ray observation details are described in Ueda~et~al. (2008) and
Akiyama~et~al. (in preparation).  The limiting fluxes are to
$1 \times 10^{-15}$ erg$^{-1}$ cm$^{-2}$ s$^{-1}$ in the soft band (0.5-2.0
keV) and $3 \times 10^{-15}$ erg$^{-1}$ cm$^{-2}$ s$^{-1}$ in the hard
band (2.0-10.0 keV). 

\subsection{Spectroscopic observations} \label{spectroscopy}

Follow-up spectroscopic observations to identify transients and  
obtain redshifts were taken during the survey in 2002. 
The follow-up was done with several ground-based  
8-10m telescopes and the ACS grism on HST and is described in Lidman~et~al. (2005), 
Morokuma~et~al. (2010) and Suzuki~et~al. (2012). Given the large number of transients, 
priority was given to transients that were likely to be SNe~Ia at z $>$ 1.
A number of factors went into computing the priority: 
the significance of the detection, the percentage increase in the brightness, the distance 
from the centre of the apparent host, the brightness of the candidate and the quality of the subtraction (the follow-up procedure is summarized in Lidman~et~al. 2005).
In total, 8 transients, half of which are beyond $z=1$, were classified as SN~Ia (see Table~3 for the references). 
In later years, additional spectroscopic observations were taken with FOCAS on  
Subaru to obtain redshifts of host galaxies after the transients had faded from view. 

\section{SN selection} \label{sn}
In order to obtain a sample of SNe~Ia we have selected variable objects
from our multi-epoch data and then applied a series of procedures to
further purify the sample. The first such procedure consists of lightcurve
fitting to reject objects whose lightcurves are inconsistent with an
SN~Ia. This procedures can be applied to all candidates. Additional
procedures make use of color, spectroscopic and additional information
available for subsamples of the full sample to further cull the
sample. These procedures are now described in detail.\\

\subsection{Variability selection} \label{selection}

The details of the initial selection of variable objects in the SXDF
are described in detail by M08.  In brief, we use
an imaging subtraction method introduced by Alard \& Lupton (1998) and
developed by Alard (2000), which enables us to match one image against
another image with a different PSF.  We can then detect and measure
variable objects in the subtracted images. This method is applied for
all possible pairs of stacked images at different epochs. 
In the subtracted images, we select objects having a flux greater than
$5\sigma_{b}$ in an aperture of 2~arcsecond in diameter, where
$\sigma_{b}$ is the background fluctuation within an aperture of this
size. A total of 1040 variable objects were detected. 
We are selecting supernovae that occurred in 2002; 
since supernovae light curves last only a few months, there
should be no variability detected in 2003 or 2005. 
Of 1040 variable objects, only 371 did not show variability 
(above $5 \sigma_{b}$) in 2003 and 2005. These 371 objects are classified as 
transients. For computing light curves, we then assume that there is no flux 
from the transient in images taken from 2003 onwards.
Finally, we require that objects show at least a $5\sigma_{b}$ 
increase in 2 or more epochs in the $i'$-band. If a variable object is only detected in one epoch,
the object might be a false detection due to galaxy missubtraction, or
another kind of transient phenomenon. Note that this requirement is
accounted for in the rate calculation during the calculation of the
control time (\S\ref{ct}). 

We applied one further consideration at this stage; at redshift $z \sim
1.4$, the central wavelength of the SuprimeCam i' band corresponds a
rest-frame wavelength of $\sim$$3250$\AA. The properties of supernovae,
of all types, are not well characterized over this wavelength region,
limiting the effectiveness of non-spectroscopic classification methods.
For this reason we remove 5 objects whose spectroscopic redshift are $z>1.4$.
However, we also note that it is still possible to detect a SN~Ia with $z > 1.4$
(see \S\ref{type2} for a possible candidate).
After this stage, 141 variable objects remain as SN candidates.\\

\subsubsection{Host galaxy redshifts} \label{redshifts}

Although we would need spectra of both host galaxies and supernovae in
order to identify host galaxies with absolute certainty, here we
simply identify the galaxy closest on the sky to each SN candidate as
its host (M08). The stacked images of SXDF are 
much deeper than the individual images, which enables us to 
detect a reliable host galaxy for every SN candidate. 
Of the 141 host galaxies, only 22 have a spectroscopic redshift.  
Redshifts for the SN host galaxies without spectroscopic redshifts 
are derived from photometric redshifts of the host galaxies 
using the multi-color photometric dataset of SXDS.
The stacked images from the Suprime-Cam observations
have depths of $B=28.4$, $V=27.8$, $R_c=27.7$, $i'=27.7$, and $z'=26.6$ 
($3\sigma$, 2~arcsecond diameter aperture; Furusawa~et~al. 2008). 
As part of the SXDS, the field
was observed by the UKIDSS/UDS survey (Warren~et~al. 2007) in the J
, H, and K bands, with respective limiting magnitudes of 24.9, 24.2, 24.6
($5\sigma$, 2~arcsecond diameter aperture). 
Additionally, the SWIRE survey (Lonsdale~et~al. 2004) obtained data in 3.6 $\mu$m and 4.5
$\mu$m bands, with respective limiting magnitudes of 23.1 and 22.4
($3\sigma$, 3.8~arcsecond diameter aperture).

Photometric redshifts are calculated using this 10 band dataset. Photometric redshift
calculations are performed using the publicly available code {\it LePhare} 
(Arnouts~et~al. 1999; Ilbert~et~al. 2006). We fit our $B$, $V$, $R_c$, $i'$ $z'$, $J$, $H$, $K$, 
$3.6 \mu m$, $4.5 \mu m$ host magnitudes with spectral energy distributions (SEDs) from PEGASE2 
(Fioc \& Rocca-Volmerange 1997, 1999) stellar population synthesis models. We use the initial mass function of 
Scalo (1986) and 15 models for star formation (SF) history. These include a constant star formation rate (SFR) 
scenario, starburst scenario, and star formation history having an exponentially decaying SFR with exponential 
time scales of $\tau_{\rm SF}=0.1$, 0.3, 0.5, 0.7, 1, 2, 3, 5, 7, 9, 10, 15, and 20 Gyrs.

\begin{figure}
\centering
\FigureFile(80mm,80mm){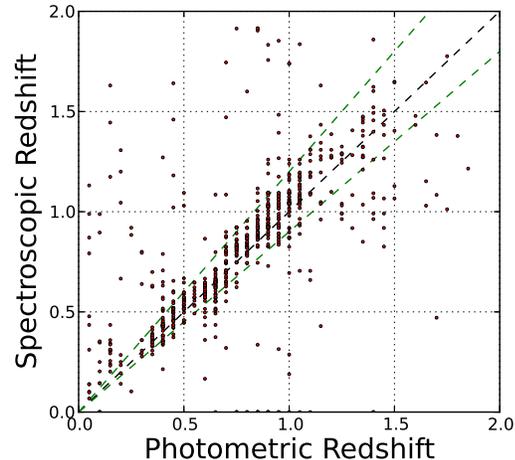}
\caption{Spectroscopic redshifts ($z_{sp}$) versus photometric
  redshifts ($z_{ph}$) of SXDS galaxies. The
  green dashed lines indicate the
  region bounding $-0.1<(z_{sp}-z_{ph})/z_{ph}<0.2$.
  \label{photz}
}
\end{figure}

To test the reliability of the photometric redshifts ($z_{ph}$), we
use 786 SXDS galaxies that have both a photometric redshift and a spectroscopic one. 
The comparison is shown in Figure~\ref{photz}. The reliability depends on the redshift range 
and the undertainty of the host-galaxy photometry. Most galaxies ($\gtrsim 80\%$) are in the
range $-0.1 < (z_{sp}-z_{ph})/z_{ph} < 0.2$. Since our work concentrates
on the determination of SN~Ia redshifts, we use the probability distribution 
function (PDF) of the host galaxy redshift when classifying its type and redshift.

It is possible that some host associations are erroneous, which       
could result in the rejection of a bona fide SN~Ia due to an errror
in the lightcurve timescale.  An erroneous association between
objects with similar redshifts --- such as those in the same group or
cluster --- is not of concern here. To check for possible host galaxy
misidentification, for the objects which we later classify as SN~Ia we
show in Figure~\ref{offset} the distribution of separations between the
SN and the center of its designated host galaxy.  According to Yasuda \&
Fukugita (2010), the radial distribution of SNe Ia is nearly consistent
with the luminosity profiles of their host galaxies.  In the case of our
SXDS candidates, a Kolmogorov-Smirnov test finds the distribution of
the SN~Ia candidates to be consistent with the luminosity profiles of
galaxies (Yasuda \& Fukugita 2010). Therefore, we assume
that erroneous host associations are unimportant for the present analysis.\\

\subsubsection{Discriminating Type~II SNe}\label{type2}

The light curves of SNe~II are
generally significantly broader than those of SNe~Ia or SNe~Ib/c. 
Therefore, as a first step we distinguish between SNe~I
(including Ia, Ib and Ic) and SNe~II using the light-curve shape.  To do
this, we compare the light curve of each candidate to template
light curves of the various SN subtypes.  As we will see, 
the light curves of most candidates are adequately sampled to distinguish 
unambiguously between SNe~I and SNe~II. Where available, we use the
spectroscopic redshift of the SN or host galaxy in the light-curve
fitting.  Where there is only a photometric redshift available, 
we use the PDF of the host galaxy calculated by {\it LePhare} as a redshift prior.

\begin{figure}
\centering
\FigureFile(80mm,60mm){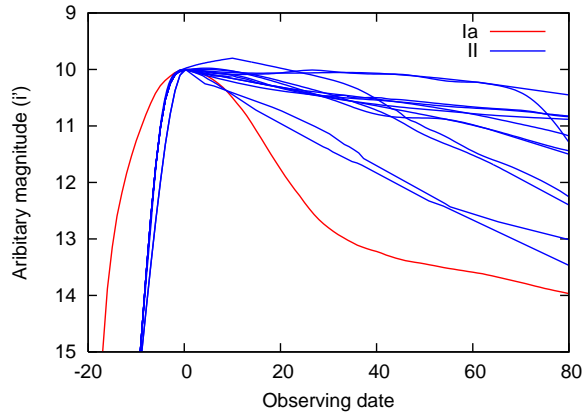}
\caption{Examples of light curve templates in the observed $i$'-band:
 A single SN~Ia template at $z=0.9$ ({\it red line}) and 12 SN~II 
 templates at $z=0.5$ ({\it blue lines}) are shown.
  \label{LCs}
}
\end{figure}

We construct SN~Ia and SN~II template light curves in the observed
$i'$-band using $K$-corrections derived from the spectral time-series
templates of Hsiao~et~al. (2007) and Nugent~et~al. (2002). The
shape of various SN~Ia light curves can be well-represented by a
single template and a stretch factor (Perlmutter~et~al. 1997).  For
the SN~Ia light curves we perform $K$-corrections with the spectral
template of Hsiao~et~al. (2007).  This template accurately describes
the UV features of SNe~Ia, which is particularly important for
high-redshift SNe~Ia.  SN~Ia light-curve shape diversity can be
neglected as it has been shown to be small compared to the difference
between SNe~Ia and SNe~II light-curve shapes (Takanashi~et~al. 2008). 
In contrast to SNe~Ia, it is impossible to describe SNe~II
with only one template.  Type~II SNe can be divided into several
subtypes (e.g., IIP, IIL and IIn) each of which exhibits a broadly
different light-curve shape. Even within subtypes there is significant
light-curve shape diversity.  Therefore, we use a set of 12
well-observed SN~II light curves as templates.  Our observed SNe~II
consist of 5 of the best-observed published SNe~II and 7 SNe~II from
the SDSS-II SN survey (Sako~et~al. 2008). In total, the SDSS-II SN
survey observed more than 50 SNe~II over three years. The 7 used here
are selected based on their discovery at an early phase and many
repeat observations ($> 10$) with long time coverage ($\sim$60 days)
in the SDSS $u'$-, $g'$-, and $r'$-bands. Details for the 12 SNe~II used
as templates are listed in Table~\ref{TypeII}. For each candidate, an $i'$-band
template light curve is made by $K$-correcting the observed multi-band
photometry to the redshift of the candidate, using the templates of
Nugent~et~al. (2002). Generally, the observed SN~II light curves lack
data points during their rising phase due to the rapid increase to
maximum after explosion. Thus, we also use the Nugent~et~al. (2002)
templates to interpolate the rising phase of the light curve.  Example
template light curves for Type~Ia and II SNe are shown in Figure~\ref{LCs}.

\begin{figure*}
\begin{center}
\FigureFile(160mm,120mm){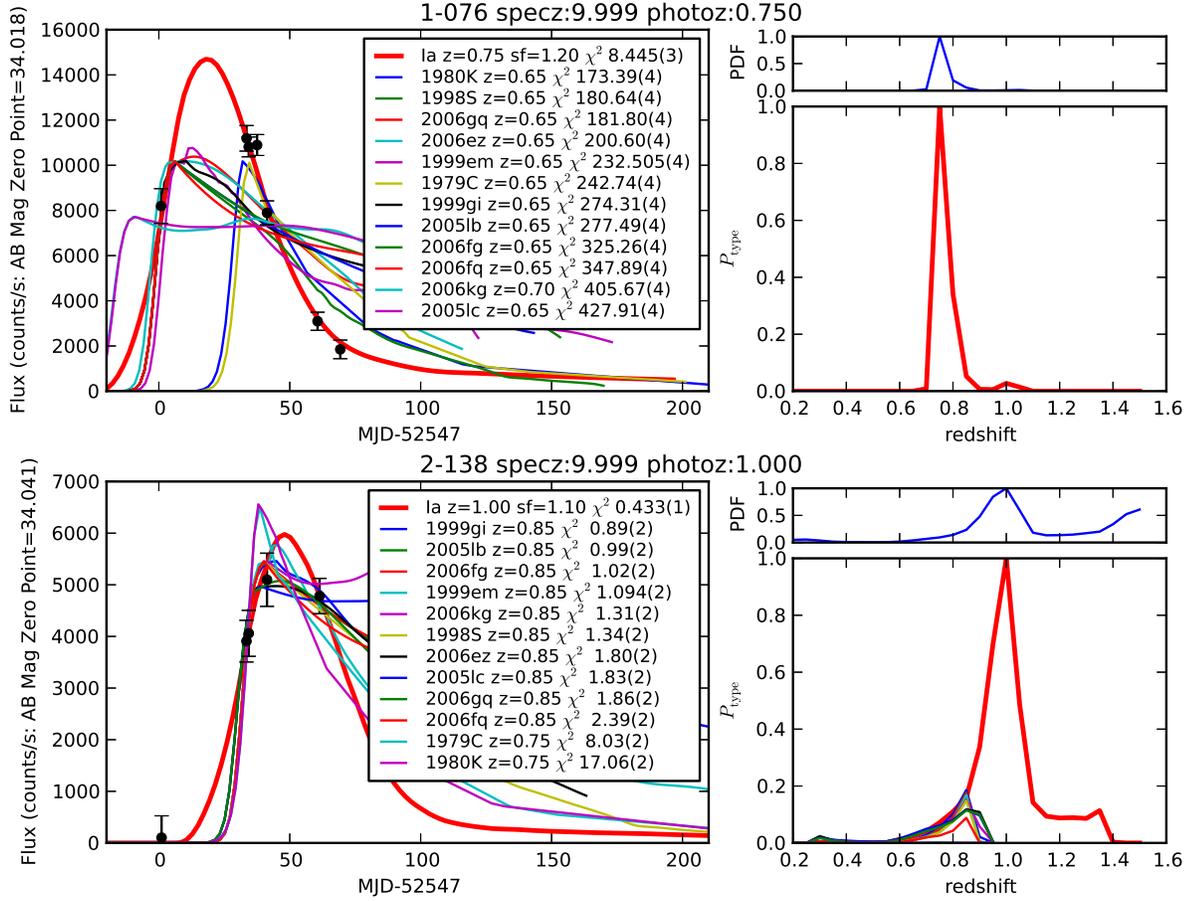}
\end{center}
\caption{Two examples demonstrating the method we use to classify SNe. 
On the left, we show the best fits using the templates used in this paper.
The numbers in the parenthesis are reduced $\chi^2$ values.
The right-hand plots show the PDFs of the host galaxies 
and the normalized $P_{type}$ as a function
of redshift. The best fit for the object in the upper panels is a SN~Ia at $z=0.75$.
This is a typical case. The object in the lower panels is an example of an 
object that has a type that is less clear. Though this object is best fit with a SN~Ia template, the 
$P_{type}$ distribution shows that SNe~II are possible. 
However, the signature of SN~Ia is still strong here and the object is classified as
a SN~Ia in our sample. The possible contamination of SN~II from the fitting 
is taken into account (see \S\ref{ratecalc}).
}
\label{LCsamples}
\end{figure*}

Using the observed $i'$-band light curve of each SXDS SN candidate, we
perform the following fitting method to refine candidates.

First, we use the probability of being a certain type of SN as a function 
of redshift using the following formula:

\begin{equation}
P_{\rm type}(z) \propto PDF(z)\times \exp \left\{-\frac{\chi^2_{\rm LC}(z)}{2}\right\}
\end{equation}

Here, $PDF(z)$ is the probability function derived by {\it LePhare}, 
and $\chi^2_{\rm LC}$  is the $\chi^2$ calculated by light curve (LC) fitting.
\begin{equation}
\chi^2_{\rm LC}(z)= \sum_{k=1}^{n}\left\{\frac{f_{obs}-f_{temp}(z)}{\Delta f_{obs}}\right\}^2,
\end{equation}
where $f_{obs}$ is the observed $i'$-band flux, $\Delta f_{obs}$ is
the observational error, $f_{temp}(z)$ is the $i'$-band flux of the
template light curve at redshift $z$, and $n$ is the number of observing epochs during 2002: 7 in 
SXDF-C and SXDF-W,  6 in SXDF-E, and 5 in SXDF-N and SXDF-S.
Note that $f_{temp}(z)$ represents a set of templates of SNe of different types.
In the light-curve fit, the free parameters on the template light curve
are the peak magnitude, the date at peak brightness, the stretch factor, and the redshift. 
Then we calculate the value of $\chi^2_{\rm LC}$ for each SN template. 
The date at peak brightness is allowed
to vary between day $-10$ and 70 where day 0 corresponds to the
beginning date of the SXDS variable object survey (September 30,
2002). The stretch factor is only used in fitting the Type~Ia
template. It is constrained to the range $0.75-1.2$ and moves
independently of peak magnitude. For SNe~Ia templates, the $B$-band
absolute magnitude is allowed to vary in the range $-20.0 < M_B <
-17.0$. This magnitude range is based on the range of the real SN~Ia
distribution observed in the SDSS-II SN survey (Dilday~et~al. 2008). 
For SNe~II templates, the $V$-band absolute magnitude is
allowed to vary in the range $-19.0 < M_V < -15.0$. As for SNe~Ia,
this range is based on the distribution of SNe~II in the SDSS-II SN
survey (see Figure~\ref{SDSSII}). In all cases, the absolute magnitude
is converted to an observed $i'$-band magnitude using the luminosity
distance and a $K$-correction with the appropriate spectral template
(Hsiao~et~al. 2007 and Nugent~et~al. 2002).  For SNe having
spectroscopic redshifts, the redshift is fixed, and $f_{temp}$ is
calculated by $K$-correcting the light-curve template to that redshift.

We determine the SN type by inspecting the value of $P_{\rm type}(z)$.
If the $P_{\rm type}(z)$ obtained by fitting the SN~Ia template is
greater than that obtained by fitting any of the SN~II templates, the candidate
is classified as a SN~I.
In order to remove candidates that are neither SNe~I nor II (i.e., AGN or 
other variable objects), we also require that the $\chi^2/d.o.f.$ for the 
best fit template be lower than 5. Based on the simulation of completeness
(\S\ref{completeness}), 98.4\% of real SNe~Ia will satisfy this requirement.

Using this method, we classify 44 of the 141 candidates as SNe~I. 
Though our control time shows a sharp drop off by $z \sim 1.4$,
one object (1-081) has been classified as SN~Ia at $z=1.45$ with $M_B=-19.53$. 
It is possible to detect a SN~Ia with $z>1.4$ if our observations cover at least two epochs 
around the maximum (see Figure~\ref{time}).
However, we do not include this object in the rate calculation and 
use only the 43 SNe~I having $z<1.4$. Examples of template light-curve fits are shown 
in Figure~\ref{LCsamples}. Although we expect this method to distinguish 
between SNe~I and II with good reliability (see following section), due to 
statistical fluctuations the classification will not be perfect. Therefore, 
we estimate and correct for completeness and SN~II contamination in \S\ref{completeness}.\\

\subsubsection{Discriminating against AGN} \label{agn}

Another potential source of contamination are AGN 
that pass our variability cuts. X-ray detection is useful in 
confirming whether or not variable objects are AGN. 
Out of 43 SN~Ia candidates, 42 objects were observed with XMM-Newton at some 
observation phase of SXDS, and only two object are detected in X-ray. 
This X-ray detection ratio is almost the same as the ratio of AGN to 
general galaxies (M08). Thus, the one object detected 
in X-ray might be supernova that occurred in a galaxy hosting an AGN. 

We have a spectrum of one of the transients associated with an X-ray
source, object 3-202 (SuF02-061). The spectrum exhibits an [Ne~III]
$\lambda\lambda 3869$ emission line, suggesting the possibility of an
AGN. [O~II]$\lambda\lambda 3727$ and H$\delta$ are also detected. From
these lines, we derive [Ne~III]/H$\delta \sim 0.35$ and
[O~II]/H$\delta \sim 0.61$, which are consistent with either an AGN or
starburst origin (Rola~et~al. 1997, P\'{e}rez-Montero 2007). 

Tests performed on the spectrum indicate that, had object 3-202 been a
SN~Ia, we would have detected the SN~Ia features.  Object 3-202 is
also very close to the core of its host galaxy, as expected for an
AGN. The offset is 0.19 pixels (0\farcs04), which is larger than our
expected measurement uncertainty.  Thus, while we are unable to rule
out the possibility of a SN~Ib/c associated with a starburst, with
this evidence we can conclusively reject the possibility that this
object is a Type~Ia SN.

We also reject object 1-143, due to the likelihood that it is an AGN. 
Unlike 3-202, we do not have a spectrum of object 1-143; however, 
it is detected in the X-rays and is closer to the core of its host galaxy
than any other candidate.

X-ray observations are usually powerful tools to detect AGN, however
not all AGN have X-ray detections and faint AGN populations are not
traced by X-ray observations (M08). Since objects
3-202 and 1-143 passed the lightcurve test but are likely AGN, we may
ask whether there are other such cases of AGN undetected by X-rays in
our sample. Only one other object, 4-203, is as close to
the core of its host as objects 3-202 and 1-143 are to theirs. 
If this was an AGN as well, our sample would have a deficit in the 
number of SNe~Ia near core, so we deem it likely that this object is not an AGN.\\

\subsubsection{Discriminating against Type~Ib/c SNe}\label{discriminatingIbc}

\begin{figure}
\centering
\FigureFile(80mm,160mm){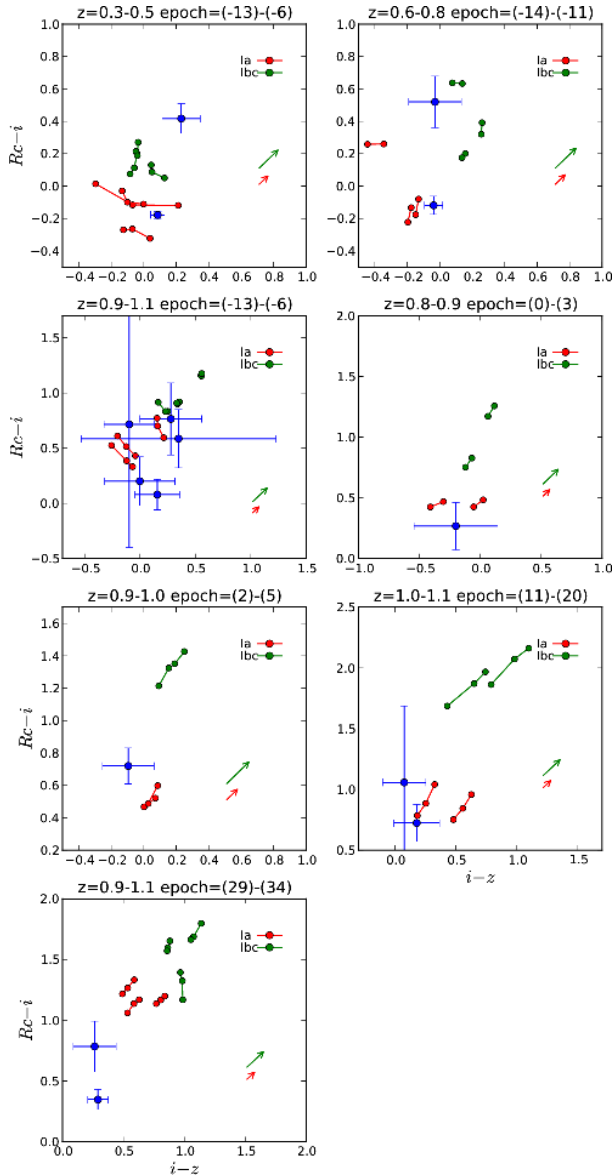}
\caption{The color-color diagrams of SN~I candidates in the observer
  frame.  The {\it red lines} and {\it green lines} indicate the
  expected colors of unreddened SNe~Ia and SNe~Ib/c respectively at
  the given redshift. The points connected by the lines indicate the
  values at specific epochs in the epoch range given.  Red and green
  arrows indicate average reddening of supernovae from their host
  galaxies: $A_B=0.4$ for SNe~Ia (Wang~et~al. 2006) and $A_B=0.7$ for
  SNe~Ib/c (Richardson~et~al. 2006).  Blue circles show the colors of
  the SN~I candidates in SXDS.  
  Out of 15 candidates, two candidates have colors most similar to SNe~Ib/c.
 \label{color}
   }
\end{figure}

At ths point we have 41 SN~Ia candidates based on their light-curve shapes.
It is impossible to further classify candidates into SNe~Ia and Ib/c 
without additional data because the shapes of SN~Ia and Ib/c
light curves are quite similar. Although SNe~Ib/c are rarer events than
SNe~Ia, we expect that the classified SNe~I will include a few SNe~Ib/c.

In our observation, some transients were observed in $R_c-$ and $z'-$bands.
We make use of this photometry to discriminate SNe~Ib/c from 
SNe~Ia based on color information ($R_c-i'$ vs $i'-z'$); 14 of the remaining SNe~I
candidates have this color information. The AGN 3-202 also has a color
measurement, as do 2 of the 8 spectroscopically-confirmed SNe~Ia. We
construct color-color diagrams of the candidates at each redshift and
epoch where the colors are available (Figure~\ref{color}).  The color
models of SNe~Ia and Ib/c are obtained from the templates of Hsiao
et~al. (2007) and Nugent et~al. (2002), respectively.  We assume an
average reddening from SN host galaxies and show the reddening as arrows
on Figure~\ref{color}. SN~Ia candidates are grouped by similar redshift and 
epoch so that we can compare the expected SN~Ia/Ibc colors and those of candidates.

Using Figure~\ref{color}, we estimate that 2 out of the 14 remaining
candidates have a color incompatible with SNe~Ia. 
These objects, 2-038 and 4-100, are therefore rejected from the sample. 
Possible contamination of SN~Ib/c on the rest of the sample is discussed in \S\ref{SNIbccontamination} 
as a systematic uncertainties.\\

\subsubsection{Properties of SN~Ia candidates} \label{properties}

We can use the small subset of the candidates that are spectroscopically confirmed 
SNe~Ia as a basic consistency check. All 8 spectroscopically confirmed SNe~Ia and 
probable SNe~Ia (Ia*) found during the SXDS are classified as 
SNe~I by the light-curve fitting.

Although the 39 SN~Ia candidates will have some contamination from SNe~II 
(estimated in \S\ref{completeness}) and SNe~Ib/c (\S\ref{SNIbccontamination}), we expect most of them 
to be SNe~Ia. We can check that most of the 39 SN~I candidates have properties 
broadly consistent with SNe~Ia. The best fit light-curve parameters for each of the 39
candidates are shown in Table~3. We note that the uncertainty in these parameters is 
often large, particularly for candidates lacking a spectroscopic redshift. 
This is not a problem however, as we are concerned with the broad light-curve characteristics 
of the sample as a whole rather than an precise determination of the light-curve parameters of any single SN. 
We discuss the distribution of absolute magnitude, light-curve width, and host galaxy separation for the candidates.\\

\begin{figure}
\FigureFile(80mm,120mm){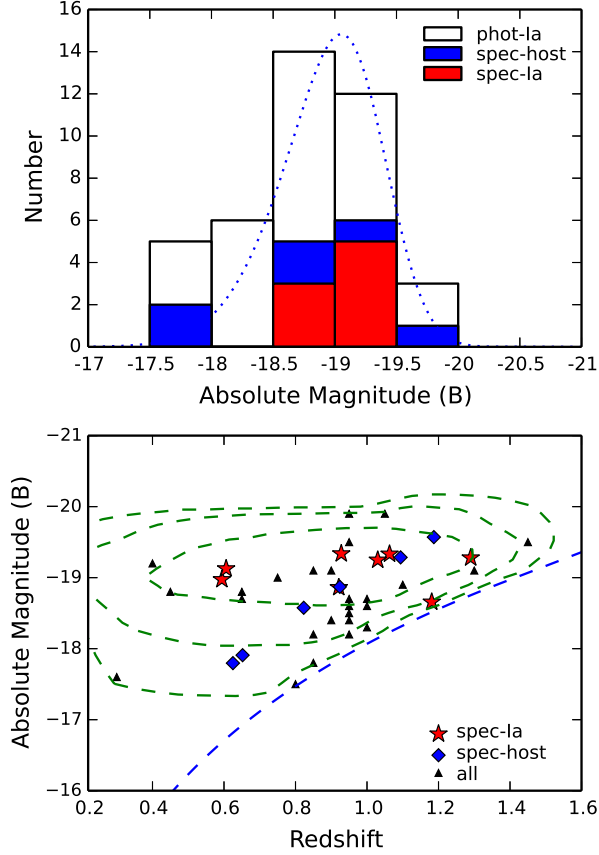}
\caption{The observed peak magnitude distribution of SNe~Ia candidates in our SXDF sample.  
  The {\it top figure} shows the number distribution. Spectroscopically-confirmed SN~Ia, 
  not spectroscopically-confirmed SN~Ia but objects with spectroscopic redshifts from host galaxies, 
  and other SN~Ia candidates are plotted as red, blue and open histograms, respectively
  The {\it bottom figure} shows the redshift distribution of the SN~Ia candidates. 
  In the bottom figure, the {\it red stars} describe the SNe~Ia SNe confirmed by spectroscopic observations,
  the {\it blue diamonds} describe the candidates with spectroscopic redshifts from their host galaxies, and 
  the {\it black triangles} describe the remaining SN~Ia candidates.
  The {\it blue dotted line} describes the limiting magnitude of the SXDS observations.
  The {\it green contours} show the 1-, 2- and 3-$\sigma$ confidence intervals 
  for the distribution of a pure SN~Ia sample calculated using the methods and SN rates from \S\ref{ratecalc}.
  } 
\label{IaSXDS}
\end{figure}

{\it Absolute magnitude}. The distribution of the candidates' $B-$band absolute magnitudes 
 (uncorrected for host galaxy extinction) is shown in Figure~\ref{IaSXDS} (top). The distribution peaks 
around $M_B \sim -19.0$, the expected average magnitude of SNe~Ia. 
The expected distribution of SN~Ia magnitudes from this survey 
(based on the simulations described in \S\ref{ratecalc}) is shown as a dotted line.
There appears to be an overabundance of faint candidates, possibly due to contamination from SNe II. 
The lower panel of Figure~\ref{IaSXDS} shows where the excess lies in redshift -- the excess faint candidates 
are found mainly at lower redshift ($z<0.8$). 
Note that the simulated expected distribution (green contours) takes into account the shift in the distribution of 
SNe toward higher luminosity and larger stretch with redshift (e.g., Howell et al. 2007). 
As a result, the top of the green contours slopes up with redshift.

{\it Light curve width}. Since our light curve fitting is based on at most seven epochs, 
constraining the light curve width (stretch parameter) is challenging
compared to other parameters, e.g., $M_B$, and the day of maximum. 
Their errors are very large ($\Delta s \sim 0.1-0.2$). Furthermore, our stretch factors in Table~3 
are not B-band stretch factors but observed $i'-$band stretch factors, 
which correspond to other rest-frame bands depending on redshift.This is not a problem here, 
however, since we are employing light curve fitting only to determine type and redshift.
The observer-frame $i'$-band stretch distribution contains a broad peak around $s \sim 1$,
which is the consistent with observations of nearby SNe~Ia. At the same time, we found that 
some faint objects have large stretches ($s \sim 1.2$), though we expect large stretches for luminous objects. 
Some of these objects might be misclassified SNe~II. We estimate the rate of misclassified SNe~II in \S\ref{misclassificationratio}.

\begin{figure}
\FigureFile(80mm,60mm){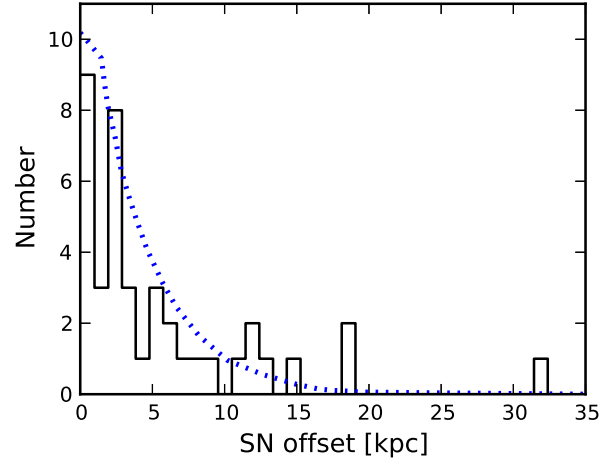}
\caption{The distribution of distance between candidate and host
  galaxy center for the SN~I candidates. The {\it black solid line}
  indicates the distribution of SN~I candidates and the {\it blue dotted
  line} indicates the luminosity profiles of galaxies (Yasuda \& Fukugita 2010).
  \label{offset}
}
\end{figure}

{\it Host galaxy separation}. We show the distribution of the distance from each candidate to the center of its 
designed host galaxy (Figure~\ref{offset}). According to Yasuda \& Fukugita (2010), the radial distribution 
of SNe~Ia is nearly consistent with the luminosity profiles of their host galaxies. 
In the case of our SXDS candidates, a Kolmogorov-Smirnov test finds the distribution of the 
SN~I candidates to be consistent with the luminosity profiles of galaxies.\\

\subsubsection{The estimated number of observed SN Ia} \label{Nestsubsection}

To count the estimated number of observed SNe~Ia ($N_{est}$), 
the easiest way might be to use the best-fit redshift derived from the fitting.
However, some SN~Ia candidates have large host photo-$z$ uncertainties 
(see bottom right figure of Fig. \ref{LCsamples} for an example).
Instead, we allocate the number of SN~Ia according to their $P_{type}$
distributions. For example, in the case of 2-138 illustrated in Fig. \ref{LCsamples}, we allocate 
a fractional contribution of 0.55 to the $0.6<z<1.0$ bin and 0.45 to the $1.0<z<1.4$ bin.
These allocations are summarized in Table~3, where it can be seen that the probability 
is concentrated in a single bin for each SN.

\section{Rate Calculation}  \label{ratecalc}

Given the supernova rate per unit comoving volume $r_V(z)$, 
the average number of SNe we expect to observe in the redshift bin
$[z_1, z_2]$ is given by
\begin{equation}
N_{exp}(z_1< z < z_2)=\int_{z_1}^{z_2}r_V(z)\frac{CT(z)}{1+z}\frac{\Theta}{4\pi}V(z)dz, \label{Rateeq}
\end{equation}
where $V(z)dz$ is the comoving volume in a redshift slice of width
$dz$, $\Theta$ is the solid angle observed in the survey (in units of steradians), 
and $CT(z)$ is the observer frame ``control time''. The control 
time can be thought of as an ``effective visibility time''; it is
the total time (in the observer frame) for which the survey is
sensitive to a SN~Ia at redshift $z$. In any survey of finite
length, the observed number of SNe in any given bin will differ from
the average expected number $N_{exp}$ due to Poisson
statistics. Given a functional form of the rate $r_V(z)$, we can
estimate its parameters by comparing the observed number of SNe to
$N_{exp}$ in each redshift bin. Alternatively, we can make the
approximation that the rate is constant within each bin. Under this
approximation, $r_V(z)$ can be moved outside the integral in
equation~(\ref{Rateeq}). Using $N_{Ia}$ as an unbiased estimator of
$N_{exp}$, we get an estimate of the rate in the bin $z_1<z<z_2$,

\begin{equation}
\widehat{r}_v(z_1<z<z_2) =
\frac{N_{Ia}(z_1<z<z_2)}{\int_{z_1}^{z_2}\frac{CT(z)}{1+z}\frac{\Theta}{4\pi}V(z)dz}.
\label{Rateeq2}
\end{equation}

Since CT(z) differs by changing position (field) and survey epochs, 
CT(z) is calculated for all fields and corresponding survey parameters, 
then normalized for the rate calculations.
In this paper, we use two methods. Assuming the rate follows a simple
power law, $r_V(z) = r_0(1+z)^\alpha$ (Pain~et~al. 2002), 
we estimate its parameters
using equation~(\ref{Rateeq}). We also use equation~(\ref{Rateeq2}) to
estimate the rate in three broad bins, $0.2 < z < 0.6$, $0.6<z<1.0$,
$1.0<z<1.4$. 

Because we have a spectroscopic classification for only a 
small minority of our SN candidates, we use a photometric typing
method as our primary means of classifying into SNe~I and II, thereby arriving at
an estimated number of SNe~Ia observed, $N_{est}$. This method can
give a biased estimate of the true number of SNe~Ia, $N_{Ia}$, 
due to the limited number and precision of observations.
Specifically, some Type~II SNe may be misclassified as SNe~Ia, while
some Type~Ia SNe may be misclassified as SNe~II. 
The estimated number of SNe~Ia, $N_{est}$, can be expressed as follows:
\begin{eqnarray}
\lefteqn{{N}_{est}(z)}\nonumber\\
&&=N_{Ia}(z)P_{Ia}(z)+N_{II}(z)F_{II}+\widetilde{N}_{Ib/c}+\widetilde{N}_{AGN}, \label{Numeq}
\end{eqnarray}where $P_{Ia}$ is the probability of correctly classifying 
a SN~Ia (completeness) and $F_{II}$ is the probability of
classifying a SN~II as a SN~Ia (contamination). 
Note that these two factors address only misclassification in the light-curve fitting, 
not uncertainties in the SN~Ia or II light-curve templates.
Those uncertainties are addressed in our estimate of the 
systematic error (\S\ref{results}). $\widetilde{N}_{Ib/c}$ and 
$\widetilde{N}_{AGN}$ are possible residual contamination from SN~Ib/c and AGN, 
as described in \S\ref{agn} and \S\ref{discriminatingIbc}.
The method we use to derive the rates is illustrated in Figure~\ref{flow}. 
In \S\ref{ct} we calculate $CT(z)$ using simulated SN~Ia light curves. In Section \S\ref{completeness}
we calculate $P_{Ia}(z)$ and $F_{II}(z)$ using simulated SN~Ia and II light curves. 
The number of SNe~II, $N_{II}(z)$, is calculated assuming the nearby SN~II rate 
and cosmic star formation history.\\

\subsection{Control time} \label{ct}

The control time is the time interval during which we can detect the
SN explosion. Here we define the time in the observer-frame. We
compute the control time as a function of redshift.\\

\subsubsection{Simulated light curves of Ia and II}\label{simulatingLC}

\begin{figure}
\centering
\FigureFile(80mm,120mm){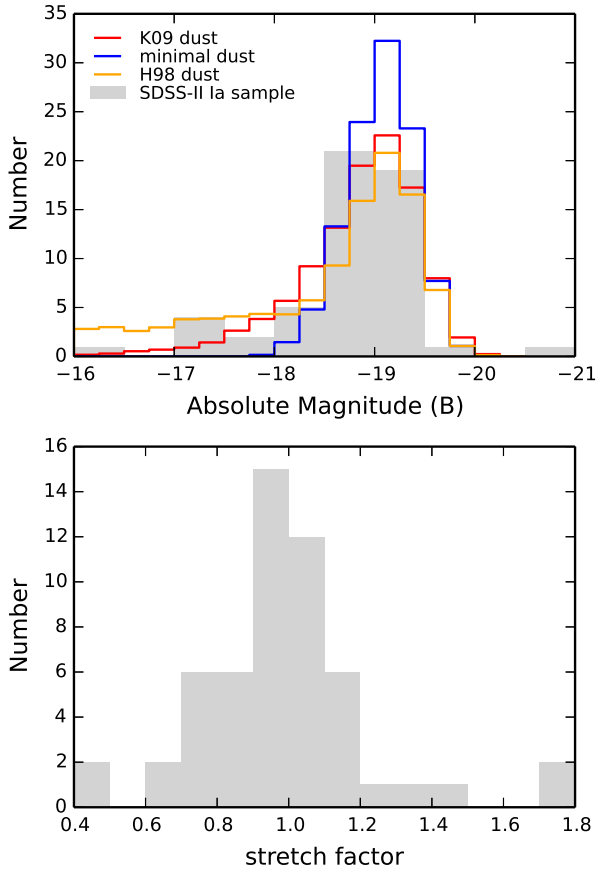} 
\caption{The absolute magnitude ({\it top}) and stretch ({\it bottom})
  distributions of the 56 $z < 0.12$ SDSS-II SNe~Ia. We make
  artificial light curves from a SN~Ia template using these
  distributions. Solid lines plotted in the top panel represent the simulated 
  SN~Ia distribution using assumptions in B12. The red line
  indicates the distribution using the dust model from Kessler~et~al. (2009),
  the blue line indicates the minimal dust model from Barbary~et~al. (2012), and the orange line 
  is the dust model of Hatano~et~al.~(1998).
  \label{IaSDSS}
}
\end{figure}

In order to calculate the control time, we carry out a Monte Carlo
simulation to generate artificial ``observed'' SN~Ia and SN~II
light curves based on the observation dates and depths of our SXDS
variable object survey. To produce a distribution of artificial SNe~Ia
modeled on the true SN~Ia distribution, we use a magnitude
distribution based on SNe~Ia from the SDSS-II SN survey (Frieman~et~al. 2008).  
As the SDSS-II sample is essentially complete at $z\leq
0.12$ (Dilday~et~al. 2008), we adopt the exact absolute magnitudes and
stretch factors of 56 $z \leq 0.12$ SDSS-II SNe for our artificial SNe~Ia.  
The 56 SNe include all spectroscopically-confirmed $z \leq 0.12$
SNe~Ia obtained in the first two years (2005 and 2006) of the survey.
The $B$-band absolute magnitude distribution (uncorrected for dust
extinction) and $B$-band stretch factor distribution of these SNe~Ia 
is shown in Figure~\ref{IaSDSS}.
The stretch distribution of SNe~Ia at high-redshift might be different 
from the local distribution; according to Howell~et~al. (2007) the average 
light curve width and average intrinsic luminosity of SNe~Ia increase 
toward high-redshift for non-subluminous SNe~Ia. 
Therefore we include the effect of the stretch evolution toward high-redshift 
in our simulation. At each redshift, we make an artificial light curve in the
observed $i'$-band based on each of the 56 SNe. To do this, we use the
absolute magnitude of the SN and a $K$-correction based on the
$u'g'r'i'$ SDSS-II light curve and the Hsiao~et~al. (2007) template.
Though we use real SNe~Ia, including whatever dust extinction they suffer, 
for our control time simulation in order to represent the actual $M_B - s$ distribution, 
an alternative approach is to employ a simple parameterized family of lightcurves, e.g., 
using stretch and color, $c$. B12 simulated the magnitude distribution of SNe~Ia using the model:
\begin{equation}\label{MBassumption}
M_B = -19.31 - \alpha (s-1) + \beta c +I
\end{equation}
where $-19.31$ is the fiducial magnitude, $\alpha =1.24$, $\beta =2.28$ 
(Kowalski~et~al. 2008), and $I$ is an additional ``intrinsic dispersion'' 
characterized with $\mu = 0.0$ mag and $\sigma = 0.15$ gaussian distribution. 
The top panel of Figure~\ref{IaSDSS} shows the comparison between simulated magnitude 
distributions for dust models from B12 and that of our sample.
Our sample is well represented by the distribution with the extinction model presented in 
Kessler~et~al. 2009 (hereafter K09), which is expressed as $P_(A_V) \propto {\rm exp}(-A_V/0.33)$ 
from host-galaxy SN extinction in the SDSS-II SN Survey. 
Among the models examined in B12, the model of Hatano~et~al. (1998), 
which was used for the main result in Dahlen~et~al. (2008), was the most dust-affected model.  
This model is also indicated in the top panel of Figure~\ref{IaSDSS}, showing tail to the fainter side.
The uncertainty caused by the choice of extinction models is discussed in \S\ref{results}.

\begin{figure}
\FigureFile(80mm,60mm){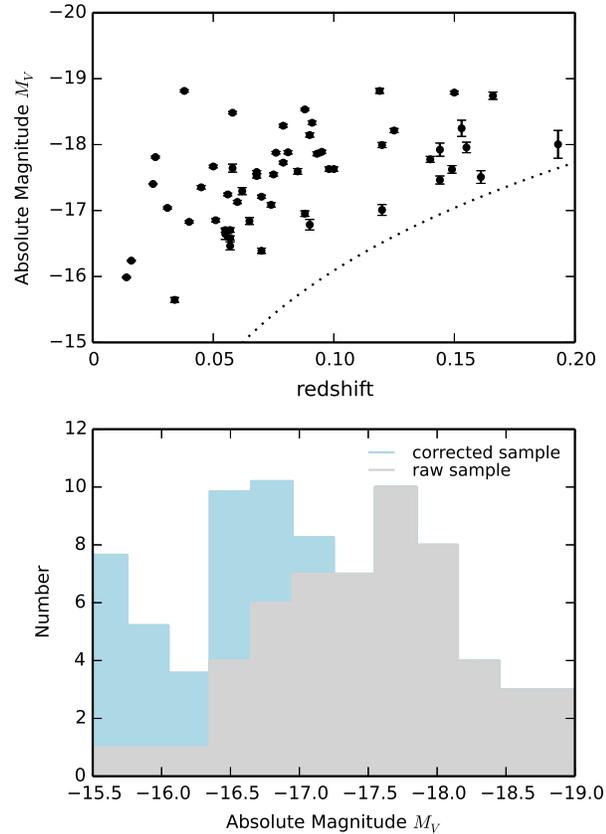}
\caption{ The absolute magnitude and redshifts of Type~II supernovae
 found by SDSS-II SN survey at $z<0.17$ ({\it top}).  
 The dashed line shows the $5\sigma$ detection limit of SDSS-II SN survey. 
 At higher redshifts fainter SNe are lost, therefore we construct a luminosity 
 function that accounts for incompleteness ({\it bottom}; see Equation~\ref{SNIIcomleteness}).
 We vary the magnitude limit curve 0.2 mag brighter/fainter to estimate systematics.
 \label{SDSSII}
   }
\end{figure}

Next, we make artificial SN~II light curves from the Type~II templates
of \S\ref{type2}.  As for SNe~Ia, the absolute magnitude distribution
for these light curves is based on real SDSS-II SNe. 
However, it is more difficult to achieve a complete sample for SNe~II because they are
intrinsically fainter than SNe~Ia on average. 
If we use the same redshift cut off as for SNe~Ia ($z = 0.12$), the number of 
faint SNe~II will be underestimated (see Figure~\ref{SDSSII}). 
To resolve this problem, we have constructed a new luminosity function that accounts for 
incompleteness using the formula below:
\begin{equation}
\displaystyle N_{eff}(L)dL = N_{sdss}(L)dL \times \frac{V_{z<0.17}}{V_{z_{max}|L}}\label{SNIIcomleteness}
\end{equation}
where $N_{SDSS}(L)dL$ is the number of SDSS SNe~II with luminosity $L-\frac{dL}{2} < L < L+\frac{dL}{2}$ 
and $V_{z_{max}|L}$ is the volume to which a SN~II can be seen above the SDSS flux limit given the luminosity, $L$.
Thus, we can simulate lightcurves using essentially all of the real SNe~II from SDSS-II.
We will estimate the systematic error due to the correction factors in \S\ref{results} 
by varying the SDSS flux limit.
In addition to a distribution in absolute magnitude, we use
two different subtypes (Type~IIP and IIL) in our generated artificial
SN~II light curves. We use a ratio of Type~II-P to Type~II-L of 2:1
(Richardson~et~al. 2002).  Also as part of our systematic error
estimate in \S\ref{results}, we vary this ratio. We generate a total of
about 100,000 SN~Ia and SN~II light curves. 
We note that statistical error and systematic error of 2\% in the 
flux is included in the light curve simulation.\\

\subsubsection{Control time calculation} \label{ctcalc}

We calculate how many days the artificial SNe~Ia can be observed. 
We add observation errors to the artificial SN~Ia light curves. 
The observing errors are calculated from the
limiting magnitudes of the SXDS observations, which includes
the effect of decreasing signal-to-noise ratio due to the subtraction of
two images.

We also include the detection efficiency in the control time calculation.
The detection efficiency of the SXDS variable object survey was
obtained by M08.  M08 estimated the detection efficiency as a function 
of magnitude in each subtraction image. 
They added artificial stars to images and detected
them in the same manner as for the real images. 
The results are shown in Figure~8 of M08.

\begin{figure}
\FigureFile(80mm,60mm){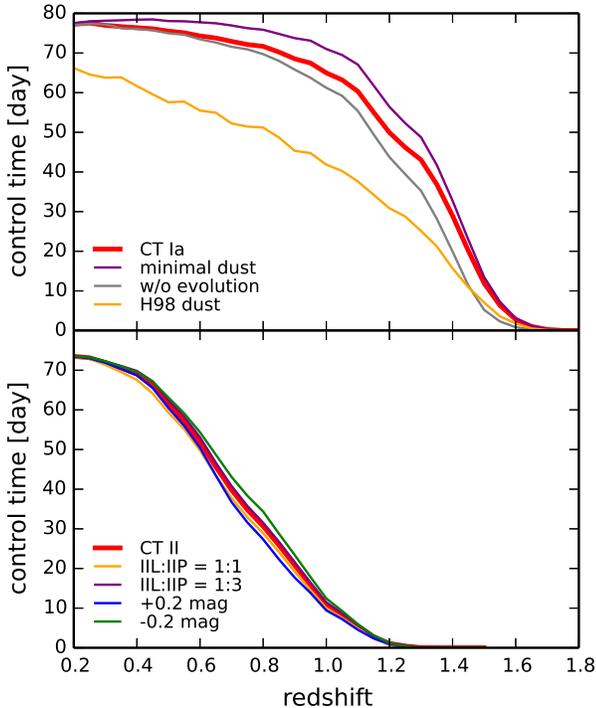}
\caption{
 {\it top}: The observer-frame control time for SNe~Ia (red line).
 The models for minimal dust (purple), without evolution effects 
 in Howell~et~al. 2007 (gray), and the model based on Hatano~et~al. 1998 (orange)
 are plotted as well. {\it bottom}: The observer-frame control time for SNe~II (red line).
 The models for different SN~IIL and SN~IIP ratio (1:1 in yellow, 1:3 in purple) are plotted.
 In addition, the choice of magnitude limit curve in Figure~\ref{SDSSII} affects the 
 luminosity function of SN~II (see the discussion in \S\ref{simulatingLC}).
 We therefore show the control time for the case of 0.2 mag brighter limit (blue) and 
 0.2 mag fainter limit (green).
 \label{time}
}
\end{figure}

The observable time duration (control time) of artificial SNe~Ia is
calculated for each redshift bin of width $\Delta z=0.05$.  The control
time of SNe~II is calculated in the same way as the SN~Ia control
time, but is only used for the estimation of contamination. 
Since the observations consist of 5 fields of Suprime-Cam with different 
survey parameters, we calculate the control times for each field,
and then weight them according to field areas (see equation 3 in B12).
In addition, we use the data from the center field when it overlaps with other fields.
The result of the weighted control time is shown in Figure~\ref{time}.\\

\subsection{Typing completeness and contamination} \label{completeness}

\subsubsection{Estimating typing completeness}\label{typingcompleteness}

\begin{figure}
\FigureFile(80mm,60mm){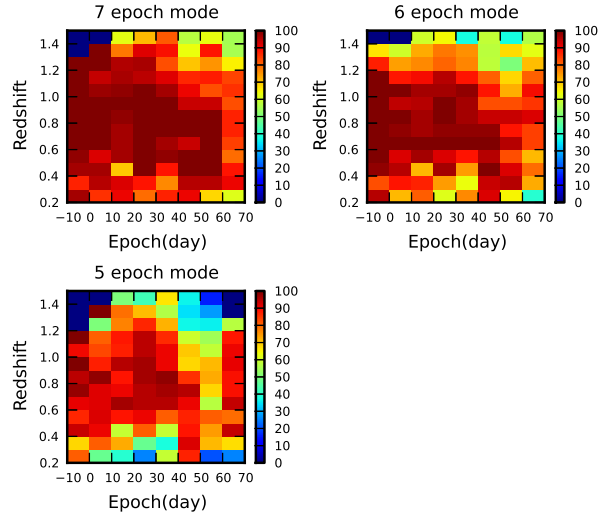}
\caption{The completeness of the light-curve fitting classification. These figures
  describe how many artificial SNe~Ia are identified as Type~Ia. The
  reliability depends on the number of observing epochs. The left top
  figure is for the 7 epoch observing mode (SXDF-C and SXDF-W). 
  The right top figure is for the 6 epoch observing mode (SXDF-E). 
  The left bottom figure is for the 5 epoch observing mode (SXDF-N and SXDF-S). 
  The horizontal axis represents the observer-frame date of the first epoch
  relative to maximum light.
  The figures show that we can classify SNe~Ia with high completeness.
 \label{comp}
}
\end{figure}

In order to determine the completeness of our light curve 
classification technique, we fit light-curve templates to our sample of artificial SNe~Ia.
Because the total number of observing epochs is
different for the different fields of the SXDF, we calculate the
completeness separately for each field. The completeness for each
field (represented as 7 epoch, 6 epoch, and 5 epoch mode) is 
shown in the different panels of Figure~\ref{comp}. 

As a general trend, the completeness improves as more epochs are observed (the second column of Table 1)
because the maximum is easily detectable. The result also shows that 
SNe~Ia are safely classified with high completeness (~80\% in average). 
Though this fraction is higher than the most efficient classification method in 
Kessler~et~al. 2010 ($\sim75$\% of Sako~et~al. 2011),
our fitting code is only used for eliminating SNe~IIL and SNe~IIP from SNe~Ia, 
whereas Kessler~et~al.~2010 attempted to eliminate SNe~Ib/c in this way.
As mentioned in \S\ref{type2}, SNe~II are more easily distinguished from SNe~Ia 
based on light curves.
Also our calculations do not include the rare luminous SNe~Ib/c
such as SN2005ap (Quimby~et~al. 2007); these will be handled as a
systematic uncertainty in \S\ref{results}.
Figure~\ref{comp} also shows that when there are fewer epochs the 
classification becomes less secure because the epoch of maximum brightness 
may be missed.\\

\subsubsection{Estimating the misclassification ratio}\label{misclassificationratio}

\begin{figure}
\FigureFile(80mm,60mm){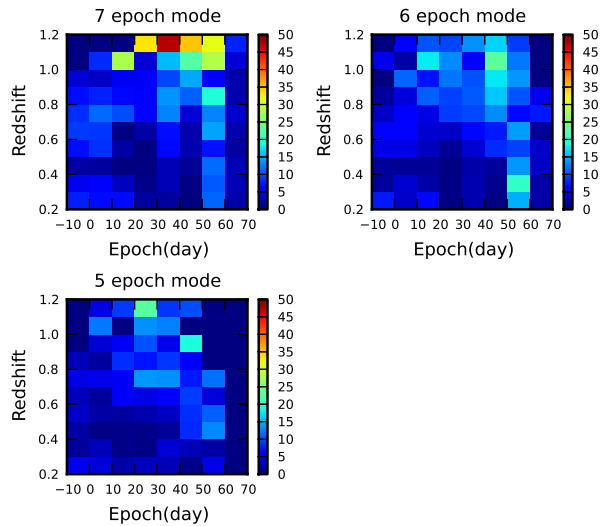}
\caption{The fraction, $F_{II}$, of instances where artificial SNe~II are 
misclassified as SNe~Ia. Three panels represent the 7 epoch, 
6 epoch, and 5 epoch mode as Figure~\ref{comp}.
The misclassification increases toward high redshift. 
In contrast, the detection efficiency of SNe~II decreases toward high redshift. 
As a result, the contamination is also small at high redshift.\\
\label{mix}
}
\end{figure}

We also estimate how often SNe~II are misclassified as SNe~Ia (the
contamination $F_{II}$). Using the same method as for SNe~Ia, we made 
artificial SNe~II and fit those SNe with light curve templates to estimate
the misclassification ratio (the ratio of SN~II classified SN~Ia).  
The results are shown in Figure~\ref{mix}. 
The misclassification becomes large toward high redshift, 
but it is not a serious problem because the detection efficiency of SNe~II 
is much less than that of SNe~Ia at high redshift (see Figure~\ref{time}).

Now, in order to find the true number ($N_{Ia}$ in Equation 5) of
SNe~Ia and II at each redshift, we need to estimate the contamination of our SNe~Ia sample by SNe~II.
There are many studies that derive core collapse supernova rates, and these results are 
consistent each other (Dahlen~et~al. 2012; Melinder~et~al. 2012; Graur~et~al. 2011;
Bazin~et~al. 2009; Botticella~et~al. 2008; Mattila~et~al. 2012; Li~et~al. 2011; Smart~et~al. 2009;
Cappellaro~et~al. 1999; Horiuchi~et~al. 2011; Magnelli~et~al. 2009).
Since higher redshift SN~II rates have larger uncertainty, we use the measured nearby 
SN~II rate and assume that the SN~II rate is increasing in proportion to the cosmic star
formation rate. The progenitors of SNe~II are massive stars and have a very short delay time between
star formation and explosion, which justifies this assumption.
We use a nearby SN~II rate of $0.394 \times 10^{-4}$ yr$^{-1}$ Mpc$^{-3}$ (Li~et~al. 2011) at $z\sim0$, 
and a star formation rate $\propto (1+z)^{3.6}$ (Hopkins \& Beacon 2006).
The SN~II rate is then $0.394 \times (1+ z )^{3.6}$.
Using this assumption, the number of SNe~II available that potentially could be misclassified as SNe~Ia, 
$N_{II}(z)$, can be calculated using the equation below.

\begin{equation}
N_{II}(z)=\int 0.394 \times (1+z)^{3.6}\frac{CT_{II}(z)}{1+z}\frac{\Theta}{4\pi}V(z)dz,
\end{equation}
where $CT_{II}$($z$) represents the control times of SNe~II.
We note in particular that this model for the SN~II rates is in agreement with the 
$z \sim 1$ SN~II rate from Dahlen~et~al. (2012).\\

\subsubsection{SN~Ib/c contamination} \label{SNIbccontamination}
In this subsection, we estimate the number of SNe~Ib/c in our sample of 39 SNe~Ia.
In addition to the 2 SNe~Ib/c that are found from their colors, we estimate that there are 
another $3.15^{+4.20}_{-2.10}$ SNe~Ib/c are in the sample. We compare this number with 
the number that one would infer from evolving local SNe Ib/c rates to higher redshift and 
we estimate the redshift distribution of the SNe Ib/c in our SN~Ia sample.

As described in \S\ref{discriminatingIbc}, the candidate refinement
indicates that approximately 2 out of the 14 SNe~I with color
information are SNe~Ib/c. This yields an estimated observed SN~Ib/c
contamination percentage of $14.3^{+18.8}_{-9.2}$ based on the combined
application of lightcurve shape and color-color selection.

Of the 39 lightcurve-selected SN~Ia candidates 8 have spectroscopic confirmation
and an additional 10 possess colors expected for SNe~Ia, leaving a
pool of 21 lightcurve-selected candidates that could still harbour
SNe~Ib/c. While it is encouraging that color classification has revealed only minor 
contamination from SNe~Ib/c, the implication is that roughly 3 additional 
SNe~Ib/c remain in this unconfirmed pool. For these a statistical correction 
can be applied if their number and redshift distribution can be estimated. 

Since spectroscopically-confirmed SNe~Ia are not present in the
unconfirmed pool, it could be argued that such objects should also be
removed from the color-color classified subsample. The rationale here
is that since spectroscopic observations may suffer greater selection
biases than color observations, the set of spectroscopically-confirmed
objects may be less representative of the unconfirmed pool than is
the set of objects with colors.  Kolmogorov-Smirnov tests of either
the redshift distributions or the peak magnitudes indicate that the
spectroscopically-confirmed and color-color subsamples are consistent with
the unconfirmed pool with $P=0.74$. While there is no direct evidence of differential
bias, this approach may be considered more conservative. In this case the
removal of 2 spectroscopically-typed SNe from the sample of 14 objects
with color-color information would raise the observed contamination rate
to 2 out of 12 objects, or $16.7^{+21.9}_{-10.8}$\%.

While correlated, these various estimates are consistent with a
SN~Ib/c contamination rate of $15^{+20}_{-10}$\%. Though the uncertainty is
large due to the small number of color samples, we adopt this rate for the
unconfirmed subsample. This would imply an additional $3.15^{+4.20}_{-2.10}$
undetected SNe~Ib/c in the unconfirmed pool. Combined with the 2
SNe~Ib/c detected via color selection this leads to a estimated
total of $5.15^{+4.20}_{-2.10}$ SNe~Ib/c in the lightcurve-shape selected sample.
This residual SN~Ib/c contamination is treated as a systematic uncertainty. 

\begin{figure}
\FigureFile(80mm,60mm){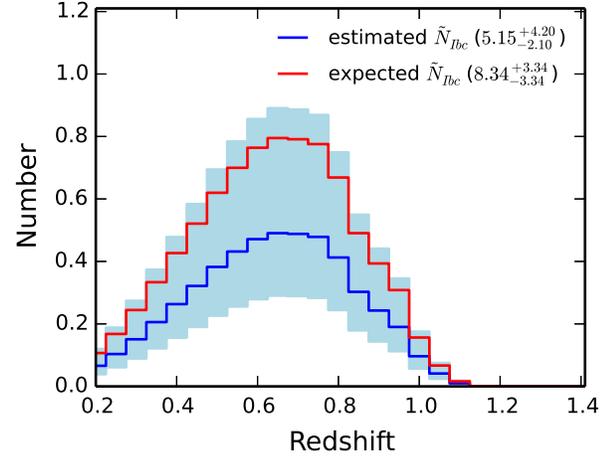}
\caption{The estimated number of possible SNe~Ib/c contamination 
  ($\widetilde{N}_{Ib/c}$ in Equation~5) in the lightcurve-selected SN~I sample.
  The {\it red solid line} indicates the expected number simulated assuming 
  nearby SN~Ib/c rate and its evolution (Li~et~al. 2011, Dahlen~et~al. 2012).
  The {\it blue solid line} and {\it light blue region} represents the number estimated 
  from color-color selection and corresponding error region.
\label{Ibcnumber}
}
\end{figure}

The second component of the statistical correction of the unconfirmed
pool is to determine the expected redshift distribution. 
The core collapse SN rate, the SN~Ib/c luminosity function and
the survey control time must then be included to predict the final
redshift distribution. Figure~\ref{Ibcnumber} shows the result of
such a calculation using the luminosity function of SN~Ib/c from nearby 
complete survey (Li~et~al. 2011) to calculate the control time, 
and assuming the observed fraction of SN~Ib/c to SN~Ia (Li~et~al. 2011) and 
core-collapse SN rate evolution (Dahlen~et~al. 2012). 
The expected number of contaminating SNe~Ib/c is $8.34^{+3.34}_{-3.34}$.
This number is consistent with the number estimated directly from the observation, $5.15^{+4.20}_{-2.10}$.
The two objects, 2-038 and 4-100, classified as a SN~Ib/c via their color are at
$z\sim0.45$ and $z\sim0.75$, respectively --- right near the peak of the 
calculated distribution in Figure~\ref{Ibcnumber}. 

In summary, we estimate that a total of $5.15^{+4.20}_{-2.10}$ SNe~Ib/c 
contaminate the lightcurve-selected SN~I sample. We use color information to 
directly remove objects 2-038 and 4-100.
The remaining $3.15^{+4.20}_{-2.10}$ SNe are subtracted using the redshift
distribution of Figure~\ref{Ibcnumber} using the appropriate scaling. 
Since we have only a limited color sample to estimate SN~Ib/c contamination, 
the impact of the uncertainty is large in the lowest bin ($0.2 < z<0.6$), 
and is comparable to the Poisson uncertainties associated with the detected 
SN~Ia sample (Table~4). However, this effect becomes much smaller 
beyond $z=0.6$, where we want to constrain the rate.
We note that reasonable changes to the color selection procedure, or to the shape of 
the SN~Ib/c redshift distribution, have negligible effect relative to the Poisson errors.\\

\section{Results and systematic error estimation} \label{results}

The SN~Ia rate is obtained from Equations~3--5. The observed 
number ($N_{Ia}$) of SNe~Ia in each redshift bin is obtained from 
Equation~5, and is shown in Figure~14. We show the SN~Ia rate in 
each redshift bin of $\Delta z = 0.4$ in Table~5. The results include 
systematic and statistical errors. These redshift bins are the same as 
those of Dahlen~et~al. (2008). The effective redshift $\bar{z}$ of each bin 
is the average redshift in the bin, weighted by the control time and volume, 
and is given by

\begin{equation}
\bar{z}=\frac{\int _{z_1}^{z_2} z \frac{CT_{Ia}(z)}{1+z}V(z)dz }{\int _{z_1}^{z_2} \frac{CT_{Ia}(z)}{1+z} V(z)dz }.  
\end{equation}

We now itemize the causes of systematic uncertainties and the methods
used to estimate their sizes, and then summarize the results in Table~\ref{sys_tab}. 

(1) In generating artificial SNe~II, we adopted a ratio of Type~II-P to Type~II-L of 2:1 
(Richardson~et~al. 2002). However the real distribution of Type~II subtypes is not 
well-known. Using a different ratio will change the fraction of misclassifications of 
SN~II as SN~Ia, $F_{II}$. 
We adopt different ratios of 1:1 and 3:1 and recalculate the SN~Ia rates in 
each case. As plotted in Figure~\ref{time}, the effect of changing the ratio is minor (up to $\sim2$\%).
The impact on the final SN~Ia rate depends on the redshift range 
(shown in Table~\ref{sys_tab}).

(2) Artificial SNe~II were generated using the peak magnitude distribution of 
low-redshift SNe~II found by the SDSS-II SN survey. 
Since the observed luminosity distribution is not complete for fainter SNe~II, 
we carefully corrected the magnitude distribution using the magnitude limit curve 
(see the discussion in \S\ref{simulatingLC}). We checked how the correction factor 
changes if the magnitude limit is moved 0.2 mag brighter/fainter.  
As in Figure~\ref{time}, the effect of changing this threshold is minor; resulting in $\sim2$\%
changes in the rate. We take the half of this difference as a systematic.

(3) In our baseline calculation we accounted for contamination from SNe~Ib/c by 
color-color selection discussed in \S\ref{discriminatingIbc}. 
This makes $3.15^{+4.20}_{-2.10}$ additional SN~Ib/c contaminating the SN~Ia sample after 
removing two objects classified as SN~Ib/c (2-038 and 4-100). 
The effect of changing the number of SN~Ib/c is greatest in the lowest-bin. 
In a bin with small number of SN~Ia, the contamination from SN~Ib/c can be the biggest uncertainty.
This means that more color information would have helped to better constrain the rate at $z<0.6$,
and such color information will be important in future surveys.

(4) In \S\ref{agn}, we removed two likely AGN. There remained one source
near core without X-ray detections or spectroscopic or color constraints. 
If that source were an AGN, or if one of the rejected
X-ray detections were a SN~Ia, the resulting error would be 2.6\%. We
take this as the systematic error due to AGN (i.e., $\widetilde{N}_{AGN}=1$ in equation~5).

(5) We separately consider the contamination from the core collapse SNe in the 
high-redshift bin ($1.0 \leq z < 1.4$). As noted in \S3.3, the control time for 
Type~II SNe is very small at $z \geq 1.0$, but there might exist very bright core collapse SNe. 
Boissier \& Prantzos (2009) show the magnitude distribution of SNe~Ib/c and II from 
a large but heterogenous sample. Most of these SNe are fainter than $\sim -18.0$ mag and there 
are very few bright SNe of $ \sim -19.0$ mag. However the magnitudes are discovery 
magnitudes, and serve only as lower limits on the peak magnitudes. 
The Ib/c sample of Richardson~et~al. (2006) contains three very bright SNe~Ic that would 
be detectable and possibly mistaken for SNe~Ia at high-redshift ($z > 1.0$). We estimate 
the contamination ratio of bright SNe using the magnitude distribution of Figure~6 of 
Bazin~et~al. (2009). There are four core collapse SNe (one object is a SN~Ib/c and three 
objects are SNe~II) that have peak absolute magnitudes that are around $-19$. 
In this figure, there are approximately 50 SNe~Ia with a similar magnitude. 
Therefore the contamination from core collapse SNe in the high-redshift bin
is estimated to be $\sim 8$\%.

(6) In the control time calculation we assumed that the average stretch 
changes with redshift in the manner prescribed by to Howell~et~al. (2007). 
If instead we recalculate the control time using the local stretch distribution 
in every redshift bin we find a somewhat lower control time in the highest 
redshift bin, leading to a $\sim 15\%$ increase in the estimated SN~Ia rate 
for $1.0 \leq z < 1.4.$. This is the dominant source of systematic uncertainty in this bin.

(7) The effect from the dust extinction of host galaxy is the most uncertain factor in the 
rate calculation. 
Recall that the sample used for the control calculation was based on 
SNe~Ia uncorrected for dust extinction. This sample has a tail extending fainter than the 
$M_B \sim -18.0$ limit given by the minimal dust model. Thus, if the minimal dust model is 
taken as a reference, our rates decrease by 2.9\%, 6.5\%, and 11.9\%, respectively, in each redshift bin.
If instead extinction is much stronger than in our observed reference sample, our rates will 
be underestimated. B12 examined various dust models to investigate the extinction effects. 
According to their estimate, the most extreme dust model based on Hatano~et~al. 1998 
(``dust model A'' in their paper) resulted in up to ~50\% changes in rates. 
Though our approach is different from B12 (i.e., they constructed SN~Ia luminosity function from
two parameter families), we examined how large the effect of this extreme dust model is.
We simulated the control time in the same manner as B12 (see Figure~\ref{time}), and estimated the 
differences to be +28.2\% , +43.0\%, and +47.8\%, respectively, in each redshift bin. 
This result is consistent with the estimate of B12. We wish to include this as a systematic 
uncertainty, but we will report it separately in the systematics in Table~4 because, 
unlike our other systematics, the size of this systematic is highly speculative.

Several studies have indicated that the mean extinction increases with redshift 
(Mannucci~et~al. 2007; Holwerda 2008), so this issue may be most important in our highest redshift bin. 
B12 took a systematic uncertainty of 50\% due to possible unaccounted for dust extinction.
As examples, in Figure~15 we show uncertainties of 50\%, and that of the minimal dust model. 
We note that even though the extinction-corrected rate is uncertain, our base measurement is 
a good representation of the rate to be used for predicting future SN searches at high redshift.

\begin{figure}
\FigureFile(80mm,60mm){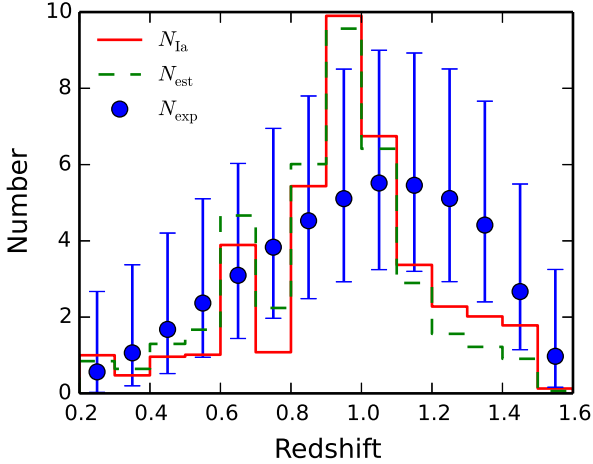}
\caption{The number of SNe~Ia from SXDS observations. The {\it green dashed 
    histogram} indicates the estimated number ($N_{est}$) of SN~Ia 
    candidates obtained by the light-curve fitting. The {\it red
    solid histogram} indicates the observed number ($N_{Ia}$) of SNe~Ia, 
   after correction for completeness and contamination.  The {\it
    blue points} with statistical error bars indicate the expected number
  ($N_{exp}$) obtained by the model-dependent rate calculation of Equation~3. 
  In this figure, one SN~Ia, 1-081, at $z=1.45$ is included in the histogram.
  \label{Num}
}
\end{figure}

\section{Discussion} \label{discussion}

\subsection{SN~Ia rate function} \label{ratefunction}

We fit the SN~Ia rates with a power law (e.g., Pain et~al. 2002),
\begin{equation}
r_V(z)=r_0(1+z)^{\alpha}. 
\end{equation}
In this fit, we use the SN~Ia rate obtained in redshift bins of
$\Delta z = 0.1$. The best fit values of $r_0$ and $\alpha$ are 
\begin{eqnarray}
r_0    &=& 0.20^{+0.52}_{-0.16}({\rm  stat.})^{+0.26}_{-0.07}({\rm syst.}) \times 10^{-4} {\rm yr}^{-1}  {\rm Mpc}^{-3} \\
\alpha &=& 2.04^{+1.84}_{-1.96}({\rm stat.})^{+2.11}_{-0.86}({\rm syst.})
\end{eqnarray}

\begin{figure}
\FigureFile(80mm,60mm){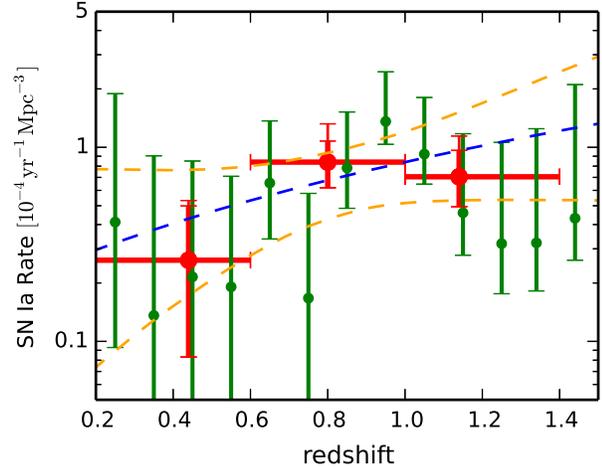}
\caption{SN~Ia rates in SXDF. {\it Red circles} show the rate in
  redshift bins of width $\Delta z=0.4$, the same bin width used by
  Dahlen~et~al. (2008), calculated using Equation~4.  
  {\it Green crosses} show the rate in redshift bins of width $\Delta z=0.1$. 
  For the upper error bar in $\Delta z=0.4$ rates, the 2nd bar represents the 
  observed + known systematics and the top bar represents the case of adding
  50\% {\it ad hoc} errors for the dust extinction.
  The {\it blue dashed line} shows $r_V(z)=r_0 (1+z)^\alpha$,
  where $r_0=0.20^{+0.52}_{-0.16}$(stat.)$^{+0.26}_{-0.07}$(syst.)$\times
   10^{-4} {\rm yr}^{-1}{\rm Mpc}^{-3}$, and  $\alpha=2.04^{+1.84}_{-1.96}$(stat.)$^{+2.11}_{-0.86}$(syst.).
  The orange dashed lines correspond to the $\pm 1\sigma$ error region.
 \label{ratefit}
}
\end{figure}

We show the fit in Figure~\ref{ratefit}, and the expected number
($N_{exp}$) obtained by this fit in Figure~\ref{Num}.

\subsection{Comparison with previous SN rate studies} \label{comparison}

\begin{figure}
\FigureFile(80mm,60mm){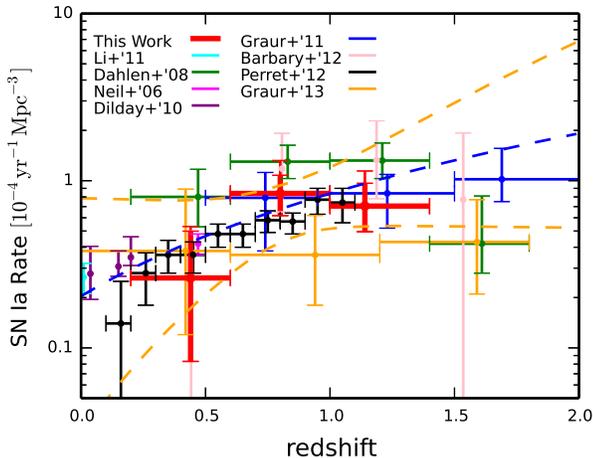}
\caption{Comparison of various previous works with our results. 
The {\it red data} show our results. As in Figure~\ref{ratefit}, the systematic uncertainty is 
divided into observed + known systematics (2nd error bar), and  
the case of adding 50\% {\it ad hoc} errors for the dust extinction (top error bar).
The {\it cyan data} are the results by Li~et~al. (2011).
The {\it green data} are the results by Dahlen~et~al. (2008). 
The {\it pink data} is the result at $z=0.47$ by Neill~et~al. (2006). 
The {\it purple data} is the result by Dilday~et~al. (2010). 
The {\it blue data} are the result of Graur~et~al. (2011).
The {\it light pink data} are the result of Barbary~et~al. (2012).
The {\it black data} are the result of Perrett~et~al. (2012).
The {\it orange data} are the result of Graur~et~al. (2013).
\label{ratepl}
}
\end{figure}

Here, we check the consistency of our results with previous works (Figure~\ref{ratepl}).
First, we compare nearby and mid-redshift SN~Ia rates with our
results. Li~et~al. (2011) derived a nearby SN~Ia rate from 274 SNe~Ia from the Lick
Observatory Supernova Search (LOSS). Dilday~et~al. (2008) measured the nearby SN~Ia 
rate at $z=0.12$ using SNe~Ia obtained by the SDSS-II SN survey.
Recently, Perrett~et~al. (2012) determined the SN~Ia rate in $0.1 \le z \le 1.1$ range 
with a very small uncertainty using the dataset of SNLS. 
These results are quite consistent with our fitted power-law curve. 
We also compare our measurements to those at similar or slightly higher redshifts.
Graur~et~al. (2011; hereafter Gr11) obtained high-redshift SN~Ia rates in the 
Subaru Deep Field (SDF) using the multi-color SED fitting classification method 
(Poznanski~et~al. 2007a). Gr11 carried out very deep photometric 
observations and detected very high-redshift SNe~Ia up to $z \sim 2.0$. 
They used one epoch with a reference epoch, and thus could not employ our light curve
fitting method.

On the other hand, Dahlen~et~al. (2008; hereafter Da08) identified SNe
with spectroscopic observations of GOODS SN survey, and measured
SN~Ia rates up to $z \sim 1.6$. B12 also derived SN~Ia
in this same redshift range from {\it Hubble Space Telescope} Cluster Supernova Survey.
Even though these studies used different samples and techniques, our result
and the results from these studies are consistent within the uncertainties. 

In the highest redshift regime ($z \gtrsim 1.4$), the SN~Ia
rate is still uncertain. Gr11 show that the SN~Ia rate remains high
beyond this redshift, while the SN~Ia rates of Da08 are flat from $z \sim 0.8$ to $z \sim
1.2$, and then show a sharp decline at $z \sim 1.6$.
As noted in \S5, B12 also found that different assumptions about
host-galaxy dust extinction can induce systematic differences
between measurements. It is therefore not yet clear if we have 
observed the peak in the volumetric SN~Ia rate.
According to galaxy studies (e.g., Hopkins \& Beacom 2006), the peak
of the cosmic star formation rate is at $z \sim 2$--3. 
Based on the peak in the cosmic star formation rate, we estimate that the delay
time of SNe~Ia at high redshifts is $\lesssim 2$-3 Gyr.  In order to
determine whether SNe~Ia with the shortest delay times are dominant or
not, it is necessary to observe more samples at $z \geq 1.4$. This
will be a key issue for future SN~Ia surveys.  One of these is the
HSC transient survey, which will use {\it Hyper Suprime-Cam} on the Subaru Telescope. 
In this survey, $\sim 100$ SNe~Ia will be detected with $z \gtrsim 1.0$, and thus provide
useful information about the high-redshift SN~Ia rates. At the same time, 
at low redshift it will construct a complete sample of SNe of all types, to faint luminosities 
and/or high values of dust extinction. Both the luminosity and extinction distributions 
will help in refining future rates calculations at high redshift.

We also check for consistency with Totani~et~al. (2008; hereafter
To08).  To08 measured the delay time distribution of SNe~Ia using the
same variable object catalog of M08, but used a
different method to select SNe~Ia.  To08 selected 65 SN~Ia candidates
showing significant spatial offset from the center of host galaxies
with an old stellar population. However, not all of their candidates
are SNe~Ia due to the uncertainty in the selection method. To08 estimated
that 82\% of the 65 candidates were actually SNe~Ia.  We check their
candidates by our light-curve fitting method.  Out of the 65 candidates
in To08, 19 candidates have light curves with
a sufficient number of epochs and sufficient
signal-to-noise to perform our light-curve fitting. Fifteen candidates
(79\% of the 19 candidates) are identified as SNe~Ia.  This result
indicates that our selection of SNe~Ia is consistent with the
selection of To08.

\section{Conclusions} \label{conclusions} 
In this paper we have presented new measurements of the 
high-redshift SN~Ia rate, using objects selected from the SXDS 
variable object survey. Each variable object was observed 
in the $i'$-band at 5-7 epochs for about two months,
which is sufficient to build light curves. The variable objects
are classified by comparing their light curves with template light curves. 
Out of 1040 variable objects, 44 SN~I candidates are selected. 
After excluding a SN~Ia beyond $z=1.4$, and using ancillary data to
exclude two likely AGNs and two likely SNe Ib/c, we construct a
sample of 39 SNe Ia that are then used to derive the rates.
Using simulated light curves, we correct the number of SN~Ia
candidates for incompleteness due to misclassifcation.
The control time is also calculated with artificial SN light curves. 
Finally, we derive the SN~Ia rate in several redshift bins between
$0.2 < z < 1.4$. Our rate measurements are the most distant yet obtained
using light curves from ground-based telescopes. 
We have considered a number of systematic factors affecting the rates.
Chief among these is the correction for extinction. However, even using conservative
estimates of the systematic error, the statistical errors are comparable in size.
Improved systematics control will be much more important for future rates measurements
based on much larger samples.

Our SNe~Ia rates are consistent with the rates that have been derived in earlier studies. 
Up to $z\sim1.4$, the rate continues to increase and is well described by a simple power law. 
The SDXS survey was relatively inefficient at finding SNe~Ia beyond $z\sim 1.4$, 
so we are unable to either confirm or refute the downturn that has
been seen in searches done with HST beyond this redshift.
The upcoming HSC transient survey is considerably more efficient in finding SNe 
beyond $z\sim 1.4$ than the SXDS survey, 
thus offering us the possibility of measuring the rates of SNe~Ia
in this important redshift range with unprecedented precision.

\bigskip

We would like to thank the Subaru Telescope staff for their invaluable assistance. 
This work was supported in part with scientific research grants 
from the Ministry of Education, Science, Culture, and Sports, Science and Technology (MEXT) of Japan,
and by the grant-in-aid for the Global COE Program "The Next Generation of Physics, Spun from Universality and Emergence" from MEXT. 
J.O, Y.I. and T.M. has been financially supported by the Japan Society for the Promotion of Science (JSPS) through the JSPS Research Fellowship.
This work was also supported in part by the Director, Office of Science,
Office of High Energy and Nuclear Physics and the Office of Advanced
Scientific Computing Research, of the U.S.  Department of Energy
(DOE) under Contract Nos. DE-FG02-92ER40704, DE-AC02-05CH11231,
DE-FG02-06ER06-04, and DE-AC02-05CH11231.

\clearpage

\begin{longtable}{lllllll}
\caption{Summary of Suprime-Cam photometry.}
\hline 
Field       & Epoch \footnotemark[$a$] & Date (UT) \footnotemark[$b$]   & $\Delta$t \footnotemark[$c$] & t$_{exp}$ [sec] & Seeing ["] \footnotemark[$d$] & m$_{lim}$ \footnotemark[$e$]\\
\hline
\hline
\endhead
\hline
\endfoot
\hline
\multicolumn{3}{l}{\hbox to 0pt{\parbox{180mm}{\footnotesize
\footnotemark[$a$] Only observations taken during 2002 contribute the number of epochs in each filter.\\
\footnotemark[$b$] Observed date in yy/mm/dd. 
When the images were stacked together, all dates for observations are included.\\
\footnotemark[$c$] Days from the first observation in each field.\\
\footnotemark[$d$] FWHM of PSF in stacked images\\
\footnotemark[$e$] Limiting magnitude of 5$\sigma$ in $2.0$ arcsec aperture.
}}}
\endlastfoot

\hline
\multicolumn{7}{c}{$i'$-band}\\
\hline
SXDF-C  &  1  &  02/09/29,30  &   0.0  &  2700  &  0.54  &  26.19  \\
SXDF-C  &  2  &  02/11/01     &  32.6  &  1860  &  0.92  &  25.76  \\
SXDF-C  &  3  &  02/11/02     &  33.5  &  1800  &  0.68  &  25.85  \\
SXDF-C  &  4  &  02/11/05     &  36.7  &  2400  &  0.70  &  26.11  \\
SXDF-C  &  5  &  02/11/09     &  40.5  &  2460  &  0.60  &  25.77  \\
SXDF-C  &  6  &  02/11/27,29  &  59.8  &  4200  &  0.72  &  26.38  \\
SXDF-C  &  7  &  02/12/07     &  68.4  &  3000  &  0.78  &  26.32  \\
SXDF-C  &  8  &  03/10/20     & 385.7  &  5760  &  1.14  &  26.53  \\
SXDF-C  &  9  &  03/10/21     & 386.5  &  7500  &  0.58  &  26.71  \\
SXDF-C  & 10  &  05/09/28     &1094.6  &  3600  &  1.00  &  26.04  \\ \hline

SXDF-N  &  1  &  02/09/29,30  &   0.0  &  3300  &  0.56  &  26.26  \\
SXDF-N  &  2  &  02/11/01     &  32.4  &  2640  &  0.96  &  25.88  \\
SXDF-N  &  3  &  02/11/02     &  33.3  &  1800  &  0.68  &  25.86  \\
SXDF-N  &  4  &  02/11/09     &  40.3  &  2100  &  0.64  &  25.78  \\
SXDF-N  &  5  &  02/11/29     &  60.3  &  3300  &  0.74  &  26.27  \\
SXDF-N  &  6  &  03/09/22     & 357.5  &  4264  &  0.60  &  26.37  \\
SXDF-N  &  7  &  03/10/02     & 367.6  &  1500  &  0.70  &  25.88  \\
SXDF-N  &  8  &  03/10/21     & 386.5  &  3000  &  0.72  &  26.14  \\ \hline

SXDF-S  &  1  &  02/09/29,30  &   0.0  &  3300  &  0.52  &  26.28  \\
SXDF-S  &  2  &  02/11/01     &  32.5  &  3600  &  1.04  &  25.91  \\
SXDF-S  &  3  &  02/11/02     &  33.4  &  1800  &  0.70  &  25.83  \\
SXDF-S  &  4  &  02/11/09     &  40.4  &  2580  &  0.66  &  25.60  \\
SXDF-S  &  5  &  02/11/29     &  60.6  &  1500  &  0.82  &  26.00  \\
SXDF-S  &  6  &  03/09/22     & 357.6  &  4500  &  0.54  &  26.45  \\
SXDF-S  &  7  &  03/10/02     & 367.7  &  2040  &  0.68  &  26.00  \\
SXDF-S  &  8  &  05/09/28     &1094.6  &  3900  &  0.96  &  26.04  \\ \hline

SXDF-E  &  1  &  02/09/29,30  &   0.0  &  3300  &  0.60  &  26.25  \\
SXDF-E  &  2  &  02/11/01     &  32.5  &  3000  &  1.04  &  25.97  \\
SXDF-E  &  3  &  02/11/02     &  33.4  &  1800  &  0.70  &  25.83  \\
SXDF-E  &  4  &  02/11/09     &  40.4  &  2820  &  0.66  &  26.78  \\
SXDF-E  &  5  &  02/11/29     &  60.6  &  1800  &  0.80  &  26.03  \\
SXDF-E  &  6  &  02/12/07     &  68.6  &  1209  &  1.54  &  25.19  \\
SXDF-E  &  7  &  03/09/22     & 357.6  &  6000  &  0.62  &  26.58  \\
SXDF-E  &  8  &  03/10/02     & 367.7  &  1271  &  0.68  &  25.71  \\ 
SXDF-E  &  9  &  03/10/21     & 386.7  &  1400  &  0.88  &  25.51  \\
SXDF-E  & 10  &  05/09/28     &1094.6  &  3600  &  0.96  &  26.11  \\ \hline

SXDF-W  &  1  &  02/09/29,30  &   0.0  &  2400  &  0.54  &  26.14  \\
SXDF-W  &  2  &  02/11/01     &  32.7  &  3000  &  0.96  &  25.96  \\
SXDF-W  &  3  &  02/11/02     &  33.5  &  1800  &  0.66  &  25.84  \\
SXDF-W  &  4  &  02/11/05     &  36.8  &  3060  &  0.76  &  25.98  \\
SXDF-W  &  5  &  02/11/09     &  40.5  &  2100  &  0.64  &  25.97  \\
SXDF-W  &  6  &  02/11/27,29  &  59.8  &  4200  &  0.74  &  26.39  \\
SXDF-W  &  7  &  02/12/07     &  68.5  &  6483  &  1.04  &  26.34  \\
SXDF-W  &  8  &  03/10/20,21  & 386.5  &  5460  &  0.66  &  26.46  \\

\hline
\multicolumn{7}{c}{R$_c$-band}\\
\hline
SXDF-N  &  1  &  02/11/01     &  32.5  &  2400  &  0.85  &  26.37  \\
SXDF-N  &  2  &  02/11/09     &  40.5  &  1920  &  0.81  &  26.02  \\
SXDF-N  &  3  &  08/01/09     &1927.3  &  2400  &  0.79  &  26.41  \\ \hline

SXDF-S  &  1  &  02/11/01     &  32.6  &  2400  &  0.83  &  26.32  \\
SXDF-S  &  2  &  02/11/09     &  40.6  &  1920  &  0.71  &  26.29  \\
SXDF-S  &  3  &  08/01/09     &1927.4  &  2400  &  0.69  &  26.52  \\ \hline

SXDF-E  &  1  &  02/11/01     &  33.6  &  2400  &  0.89  &  26.26  \\
SXDF-E  &  2  &  02/11/09     &  40.6  &  1920  &  0.87  &  26.12  \\
SXDF-E  &  3  &  08/01/09     &1927.4  &  2400  &  0.73  &  26.46  \\ \hline

SXDF-W  &  1  &  02/11/01     &  33.7  &  3360  &  0.91  &  26.27  \\
SXDF-W  &  2  &  02/11/09     &  40.7  &  1920  &  0.65  &  25.92  \\
SXDF-W  &  3  &  08/01/09     &1927.5  &  2400  &  0.81  &  26.33  \\

\hline
\multicolumn{7}{c}{$z'$-band}\\
\hline
SXDF-N  &  1  &  02/11/04     &  35.3  & 2400 &  0.81 & 25.16 \\
SXDF-N  &  2  &  02/11/05     &  36.2  &  600 &  0.79 & 24.22 \\
SXDF-N  &  3  &  02/11/10     &  41.3  & 2340 &  1.03 & 24.93 \\
SXDF-N  &  4  &  03/09/27     & 362.4  & 4860 &  0.71 & 25.99 \\ \hline

SXDF-S  &  1  &  02/11/04     &  35.4  & 2400 &  0.77 & 25.11 \\
SXDF-S  &  2  &  02/11/05     &  36.3  &  600 &  0.81 & 24.18 \\
SXDF-S  &  3  &  02/11/10     &  41.4  & 2700 &  0.85 & 24.97 \\
SXDF-S  &  4  &  03/09/22     & 357.7  & 4800 &  0.69 & 25.61 \\ \hline

SXDF-E  &  1  &  02/11/04     &  35.4  & 2280 &  0.71 & 25.26 \\
SXDF-E  &  2  &  02/11/05     &  36.3  & 1200 &  1.07 & 24.42 \\
SXDF-E  &  3  &  02/11/10     &  41.4  & 4440 &  0.85 & 25.32 \\
SXDF-E  &  4  &  03/09/26     & 361.5  & 4800 &  0.69 & 25.81 \\ \hline

SXDF-W  &  1  &  02/11/04     &  35.6  & 2700 &  0.81 & 25.19 \\
SXDF-W  &  2  &  02/11/05     &  36.4  &  750 &  0.75 & 24.37 \\
SXDF-W  &  3  &  02/11/10     &  41.6  & 6210 &  1.05 & 25.41 \\
SXDF-W  &  4  &  03/09/27     & 362.4  & 4800 &  0.69 & 25.92 \\
\end{longtable}

 \begin{table*}
 \begin{center}
 \caption{Type~II supernova templates}
 \begin{tabular}{llcc}
 \hline
 Name  & Reference & Type  &  Lightcurve points \\
 \hline
 \hline
SN1979C  & DeVaucouleurs~et~al. 1981\footnotemark[$a$]  & II-L & 23 (U), 32 (B), 31(V)\\
SN1980K  & Buta 1982\footnotemark[$b$]  & II-L & 20 (U), 25 (B), 25 (V)\\
SN1998S  & Fassia~et~al. 1998\footnotemark[$c$]  & II-n & 25 (B), 26 (V), 28 (R), 21 (I) \\
SN1999em & Leonard~et~al. 2002\footnotemark[$d$] & II-P & 29 (U), 40 (B), 41 (V), 46 (R), 44 (I) \\
SN1999gi & Leonard~et~al. 2002\footnotemark[$e$] & II-P & 29 (B), 30 (V), 30 (R), 30 (I) \\
SN2005lb & Sako~et~al. in prep.\footnotemark[$f$]  & II   & 1 ($u$'), 15 ($g$'), 14 ($r$'), 12 ($i$'), 13 ($z$') \\
SN2005lc & Sako~et~al. in prep.\footnotemark[$f$] & II   & 20 ($u$'), 24 ($g$'), 24 ($r$'), 23 ($i$'), 25 ($z$') \\
SN2006ez & Sako~et~al. in prep.\footnotemark[$f$] & II   & 21 ($u$'), 21 ($g$'), 22 ($r$'), 20 ($i$'), 22 ($z$') \\
SN2006fg & D'Andrea~et~al. 2009\footnotemark[$f$] & II   & 13 ($u$'), 17 ($g$'), 16 ($r$'), 17 ($i$'), 19 ($z$') \\ 
SN2006fq & D'Andrea~et~al. 2009\footnotemark[$f$] & II   & 12 ($u$'), 22 ($g$'), 22 ($r$'), 21 ($i$'), 22 ($z$') \\
SN2006gq & D'Andrea~et~al. 2009\footnotemark[$f$] & II   & 0 ($u$'), 18 ($g$'), 16 ($r$'), 16 ($i$'), 13 ($z$') \\
SN2006kg & Sako~et~al. in prep.\footnotemark[$f$] & II   & 20 ($u$'), 20 ($g$'), 20 ($r$'), 20 ($i$'), 19 ($z$') \\
 \hline
\multicolumn{4}{@{}l@{}}{\hbox to 0pt{\parbox{180mm}{\footnotesize
 \par\noindent
 \footnotemark[$a$] 76-,91-,205-cm telescopes of the McDonald Observatory
 \par\noindent
 \footnotemark[$b$] 76-,91-cm telescopes of McDonald Observatory and 1.55-m telescope of the U.S. Naval Observatory
 \par\noindent
 \footnotemark[$c$] The 82-cm Instituto de AstrofoAsica de Canarias telescope (IAC80) on Tenerife, the 1.0-m Jacobus\\ 
Kapteyn Telescope (JKT) on La Palma and the 3.5-m Wisconsin-Indiana-Yale-NOAO telescope (WIYN) at Kitt Peak
 \par\noindent
 \footnotemark[$d$] 76-,91-,205-cm telescopes of the McDonald Observatory
  \par\noindent
 \footnotemark[$e$] The Katzman Automatic Imaging Telescope (KAIT) and the 1.2-m telescope at the Fred Lawrence Whipple\\
Observatory (FLWO).
  \par\noindent
 \footnotemark[$f$] The SDSS's 2.5-meter telescope at Apache Point Observatory
    }\hss}}
  \end{tabular}\label{TypeII}
\end{center}
\end{table*}

\clearpage

\thispagestyle{plain}
\voffset 1.0cm

\begin{table}[htbp]
\rotatebox{90}{\begin{minipage}{\textheight}
\centering
\caption{SN~Ia candidates}
\begin{tabular}{llccccccccccccccc}
\hline
&&&&&&&&& \multicolumn{4}{c}{The allocation of SN~Ia to each bin}\footnotemark[$d$] &&&&\\
Field-ID & SuF02-ID\footnotemark[$a$] & Offset\footnotemark[$b$] & RA & Dec & Host ID &
$z_{phot}$ & $z_{spec}$ & $z_{SN}$\footnotemark[$c$] & $_{0.2-0.6}$& $_{0.6-1.0}$  & $_{1.0-1.4}$  & $_{1.4-}$ &  $M_B$\footnotemark[$e$] & $m_{i'}$ & $s$\footnotemark[$f$] & Day\footnotemark[$g$]\\
\hline
\hline
1-018 & SuF02-007  & 11.944 & 02:18:52.178 & -05:01:13.17 & iC-083788 & 0.85 & - & 0.950 & 0.028 & 0.650 & 0.313 & 0.009  & -18.5 & 24.7 & 1.15 & 32\\
1-020 &  SuF02-012\footnotemark[$h2,3$] & 0.205 & 02:18:51.576 & -04:47:25.97 & iC-176827 & 1.15 & 1.290 & 1.290 & - & - & 1.000 & - & -19.3 & 25.1 & 1.15 & 34\\
1-038 & SuF02-J02  & 2.490 & 02:18:42.854 & -05:04:12.50 & iC-063533 & 0.70 & 0.652 & 0.652 & - & 1.000 & - & - & -17.9 & 24.5 & 1.10 & 36\\
1-045 & SuF02-082  & 4.756 & 02:18:40.690 & -05:03:43.73 & iC-065953 & 0.65 & 0.625 & 0.625 & - & 1.000 & - & - & -17.8 & 24.5 & 1.20 & 52\\
1-076 &   & 7.785 & 02:18:31.195 & -05:01:23.60 & iC-081825 & 0.75 & - & 0.750 & - & 0.985 & 0.015 & - & -19.0 & 23.6 & 1.20 & 18\\
1-090 &   & 3.360 & 02:18:22.051 & -04:57:18.78 & iC-107586 & 0.85 & - & 0.850 & - & 1.000 & - & - & -18.2 & 24.7 & 1.20 & 14\\
1-120 & SuF02-004  & 0.378 & 02:18:09.026 & -04:54:18.71 & iC-126749 & 1.00 & 1.187 & 1.187 & - & - & 1.000 & - & -19.6 & 24.6 & 1.20 & 30\\
1-157 &   & 2.550 & 02:17:50.203 & -05:03:45.62 & iC-066598 & 0.45 & - & 0.450 & 0.993 & 0.007 & - & - & -18.8 & 22.8 & 0.75 & 70\\
1-175 & SuF02-000\footnotemark[$h3$]  & 2.953 & 02:17:42.590 & -05:06:33.76 & iC-049015 & 0.75 & 0.921 & 0.921 & - & 1.000 & - & - & -18.9 & 24.2 & 1.15 & 44\\
1-192 & SuF02-065\footnotemark[$h1$]  & 1.800 & 02:17:34.589 & -05:00:16.78 & iC-088216 & 0.58 & 1.181 & 1.181 & - & - & 1.000 & - & -18.7 & 25.5 & 1.20 & 38\\
1-193 & SuF02-060\footnotemark[$h1$] & 1.044 & 02:17:34.565 & -04:53:47.36 & iC-129501 & 1.05 & 1.063 & 1.063 & - & - & 1.000 & - & -19.3 & 24.2 & 1.00 & 44\\
1-202 &   & 1.496 & 02:17:32.114 & -04:53:30.42 & iC-131867 & 1.05 & - & 1.100 & - & 0.094 & 0.906 & - & -18.9 & 24.9 & 0.75 & 52\\
1-203 &   & 5.611 & 02:17:32.160 & -05:11:15.69 & iC-017504 & 0.90 & - & 0.900 & - & 0.998 & 0.002 & - & -19.1 & 24.0 & 1.10 & 0\\
1-242 & SuF02-002  & 2.535 & 02:17:12.264 & -04:55:08.82 & iC-121417 & 0.75 & 0.823 & 0.823 & - & 1.000 & - & - & -18.6 & 24.2 & 1.00 & 36\\
1-252 &   & 1.502 & 02:17:10.188 & -04:50:43.91 & iC-148634 & 0.95 & - & 0.950 & - & 0.945 & 0.054 & -& -18.4 & 24.8 & 1.20 & 60\\
1-254 &   & 0.608 & 02:17:09.758 & -04:57:47.89 & iC-105194 & 0.95 & - & 0.950 & - & 0.906 & 0.094 & - & -19.5 & 23.7 & 0.95 & 8\\
1-258 & SuF02-071\footnotemark[$h3$]  & 8.003 & 02:17:08.772 & -05:02:06.24 & iC-076434 & 0.90 & 0.928 & 0.928 & - & 1.000 & - & - & -19.3 & 23.8 & 0.95 & 36\\
1-280 & SuF02-027\footnotemark[$h3$]  & 2.005 & 02:17:00.074 & -04:58:20.14 & iC-100962 & 0.60 & 0.594 & 0.594 & 1.000 & - & - & - & -19.0 & 23.2 & 1.00 & 38\\
2-019 & SuF02-026  & 1.720 & 02:18:51.905 & -04:46:57.29 & iN-013001 & 0.35 & - & 0.800 & 0.013 & 0.969 & 0.015 & 0.003 & -17.5 & 25.2 & 0.75 & 24\\
2-033\footnotemark[$j$] &   & 17.026 & 02:18:42.509 & -04:34:17.99 & iN-090581 & 0.35 & - & 0.400 & 1.000 & - & - & - & -19.2 & 22.1 & 0.90 & 54\\
2-042 &   & 4.048 & 02:18:31.850 & -04:25:13.52 & iN-149479 & 0.65 & - & 0.650 & 0.416 & 0.584 & - & - & -18.7 & 23.7 & 1.15 & -4\\
2-138\footnotemark[$j$]  &   & 0.752 & 02:17:46.010 & -04:36:46.59 & iN-077467 & 1.00 & - & 1.000 & 0.002 & 0.532 & 0.464 & - & -18.7 & 24.7 & 1.20 & 48\\
2-146\footnotemark[$j$]  & SuF02-037  & 1.444 & 02:17:43.363 & -04:30:57.22 & iN-113117 & 0.85 & 0.924 & 0.924 & - & 1.000 & - & - & -18.9 & 24.3 & 1.10 & 38\\
2-167\footnotemark[$j$]  & SuF02-05  & 8.014 & 02:17:27.401 & -04:40:45.42 & iN-052480 & 0.85 & - & 0.850 & - & 0.970 & 0.021 & 0.009 & -17.8 & 25.1 & 0.90 & 38\\
2-175 &   & 0.250 & 02:17:18.859 & -04:30:26.53 & iN-116245 & 1.00 & - & 1.000 & - & 0.243 & 0.757 & - & -18.3 & 25.1 & 1.15 & 14\\
3-008\footnotemark[$j$] &   & 0.294 & 02:19:04.080 & -05:14:50.16 & iS-162495 & 0.90 & - & 0.900 & - & 0.995 & 0.005 & - & -18.4 & 24.7 & 1.20 & 56\\
3-156 &   & 4.336 & 02:17:48.098 & -05:27:44.54 & iS-074610 & 0.65 & - & 0.650 & 0.004 & 0.996 & - & - & -18.8 & 23.6 & 1.15 & 66\\
4-083 &   & 11.480 & 02:20:13.834 & -05:07:34.27 & iE-135660 & 1.30 & - & 1.300 & - & 0.000 & 0.996 & 0.004 & -19.1 & 25.3 & 1.05 & 38\\
4-105\footnotemark[$j$]  &   & 3.613 & 02:20:05.904 & -05:05:31.83 & iE-123124 & 0.95 & - & 0.950 & - & 0.608 & 0.281 & 0.111 & -18.2 & 25.0 & 0.90 & 52\\
4-106\footnotemark[$j$] &   & 4.766 & 02:19:26.021 & -05:05:06.94 & iE-059305 & 1.00 & - & 0.950 & - & 0.747 & 0.253 & - & -18.7 & 24.5 & 0.75 & 54\\
4-117\footnotemark[$j$]  &   & 1.345 & 02:20:16.889 & -05:03:50.67 & iE-141288 & 0.95 & - & 0.950 & - & 0.999 & 0.001 & - & -19.9 & 23.1 & 1.20 & 8\\
4-150 &   & 13.697 & 02:19:35.458 & -04:57:42.26 & iE-076200 & 3.00 & - & 0.300 & 1.000 & - & - & - & -17.6 & 23.1 & 0.75 & -6\\
4-174\footnotemark[$j$]  &   & 0.538 & 02:19:10.296 & -04:54:19.19 & iE-036303 & 1.10 & - & 1.050 & - & 0.006 & 0.994 & - & -19.9 & 23.6 & 0.95 & 12\\
4-203 &   & 0.024 & 02:20:32.179 & -04:49:58.86 & iE-157248 & 0.95 & - & 0.950 & - & 0.666 & 0.334 & - & -18.6 & 24.6 & 1.10 & 64\\
4-233 &   & 0.415 & 02:19:49.937 & -04:45:53.30 & iE-097735 & 0.80 & - & 0.850 & - & 1.000 & - & - & -19.1 & 23.7 & 0.75 & 70\\
5-029\footnotemark[$j$]  & SuF02-025\footnotemark[$h1$]  & 0.722 & 02:16:23.947 & -04:49:29.51 & iW-069337 & 0.55 & 0.606 & 0.606 & - & 1.000 & - & - & -19.1 & 23.1 & 1.05 & 54\\
5-035 &   & 1.244 & 02:16:46.918 & -04:51:21.98 & iW-036122 & 1.00 & - & 1.000 & - & 0.579 & 0.421 & - & -18.6 & 24.8 & 0.85 & 62\\
5-049\footnotemark[$j$]  &   & 5.846 & 02:16:50.143 & -04:54:33.53 & iW-032007 & 1.00 & 1.094 & 1.094 & - & - & 1.000 & - & -19.3 & 24.5 & 0.80 & 54\\
5-149\footnotemark[$j$]  & SuF02-017\footnotemark[$h1$]  & 20.365 & 02:16:45.530 & -05:09:48.28 & iW-038187 & 0.50 & 1.030 & 1.030 & - & - & 1.000 & - & -18.9 & 24.5 & 1.00 & 30\\
\hline
\end{tabular}
\end{minipage}
}
\end{table}

\clearpage

\setcounter{table}{2}
\begin{table}[htbp]
\rotatebox{90}{\begin{minipage}{\textheight}
\centering
\caption{(Continued.)}
\begin{tabular}{llccccccccccc}
\hline
Field-ID & SuF02-ID\footnotemark[$a$] & Offset\footnotemark[$b$] & RA & Dec & Host ID &
$z_{phot}$ & $z_{spec}$ & $z_{SN}$\footnotemark[$c$] & $M_B$\footnotemark[$e$] & $m_{i'}$ &
$s$\footnotemark[$f$] & Day\footnotemark[$g$]\\
\hline
\hline
\multicolumn{13}{c}{Candidates rejected as AGN or SN Ib/c, and one possible SN~Ia at $z>1.4$}\\
\hline
1-081\footnotemark[$l$] &   & 1.134 & 02:18:29.273 & -05:01:05.60 & iC-083570 & 1.450 & - & 1.450 & -19.5 & 25.3 & 0.95 & 62\\
1-143\footnotemark[$i$]  & SuF02-058  & 0.047  & 02:17:59.674 & -04:52:26.77 & iC-138418 & 2.600 & - & 0.50  & -17.7  & 24.1 & 1.20 & 56\\
2-038\footnotemark[$k$] & SuF02-077  & 2.736 & 02:18:35.210 & -04:26:39.90 & iN-140425 & 0.450 & - & 0.450 & -17.6 & 24.0 & 0.90 & 48\\
3-202\footnotemark[$i,j$] & SuF02-061  & 0.192 & 02:17:22.769 & -05:16:56.58 & iS-148873 & 0.700 & 1.085 & 1.100 & -19.7 & 24.0 & 0.85 & 22\\
4-100\footnotemark[$k$] &   & 0.347 & 02:19:56.690 & -05:05:55.86 & iE-108255 & 0.750 & - & 0.750 & -18.2 & 24.3 & 1.10 & 54\\
\hline
\multicolumn{8}{@{}l@{}}{\hbox to 0pt{\parbox{180mm}{\footnotesize
 \footnotemark[$a$] The IDs correspond to the spectroscopic candidates in the other papers (see Lidman~et~al. 2005, Morokuma~et~al. 2010, and Suzuki~et~al. 2012).\\
 \footnotemark[$b$] The offset means distance (pixel) from objects to the center of the host galaxy.\\
 \footnotemark[$c$] The redshifts are estimated by the light curve fitting with photometric redshifts of host galaxies. When available, the spectroscopic redshifts are used.\\
 \footnotemark[$d$] The allocation of SN~Ia to $0.2<z<0.6$, $0.6<z<1.0$, $1.0<z<1.4$, and $1.4$ bins according to $PDF(z)$ (see \S\ref{Nestsubsection})\\
 \footnotemark[$e$] The absolute magnitudes are not corrected for extinction from SN host galaxies.\\
 \footnotemark[$f$] The stretch factors are not $B$-band stretch but just observed $i'$-band stretch used in the light curve fitting.\\
 \footnotemark[$g$] The days means the dates from the peak magnitude.\\
 \footnotemark[$h$] Spectroscopically confirmed SNe~Ia (including probable Ia). The numbers represent the reference: 1 is published in Lidman~et~al.~(2005), 2 is published in Morokuma~et~al.~(2010), and 3 is going to be included in Rubin~et~al.~(in prep.)\\
 \footnotemark[$i$] Detected in X-rays with XMM-Newton.\\
 \footnotemark[$j$] Candidates with color information (detected in $R_c$, $i'$ and $z'$) \\
 \footnotemark[$k$] Failed color cut.\\
 \footnotemark[$l$] This object at $z=1.45$ is not included in the rate calculation.\\
    }\hss}}
\end{tabular}
\end{minipage}
}
\end{table}

\clearpage

 \begin{table*}
 \begin{center}
 \caption{Summary of systematic uncertainties in percent}
 \begin{tabular}{llccc}
 \hline
 Source items &  num./denom.$^a$ & $0.2\leq z<0.6$ & $0.6\leq z<1.0$ & $1.0\leq z<1.4$\\[0.5 ex]
 \hline
 \hline
(1) Ratio of IIP and IIL  & numerator & $^{+0.6}_{-0.3}$ & $^{+0.8}_{-0.4}$ & $^{+0.8}_{-0.4}$ \\[1.0 ex]
(2) Luminosity function of SN~II & numerator & $^{+0.3}_{-0.6}$ & $^{+1.4}_{-1.4}$ & $^{+1.1}_{-0.9}$ \\[1.0 ex]
(3) Discrimination of Type~Ib/c  &  numerator & $^{+22.6}_{-45.5}$ & $^{+6.2}_{-12.4}$ & $^{+0.4}_{-0.8}$ \\[1.0 ex]
(4) AGN contamination   &  numerator & $^{+0.0}_{-2.6}$ & $^{+0.0}_{-2.6}$ & $^{+0.0}_{-2.6}$ \\[1.0 ex]
(5) CC SN contamination at high-z  & numerator & -                 & -                  & $^{+0.0}_{-8.0}$ \\[1.0 ex]
(6) Evolution effect of SNe~Ia  & denominator & $^{+0.5}_{-0.0}$   & $^{+3.2}_{-0.0}$ & $^{+14.5}_{-0.0}$\\[1.0 ex]
(7) Extinction (known) &  denominator  & $^{+0.0}_{-2.9}$ & $^{+0.0}_{-6.5}$ & $^{+0.0}_{-11.9}$\\[1.0 ex]
\hline
\multicolumn{2}{l}{Total known systematics} & $^{+22.6}_{-45.7}$ & $^{+7.1}_{-14.3}$ & $^{+14.5}_{-14.7}$ \\[1.0 ex]
\hline
(7)' Extinction (unknown)  &  denominator & $^{+50.0}_{-0.0}$ & $^{+50.0}_{-0.0}$ & $^{+50.0}_{-0.0}$\\[1.0 ex]
\hline
\multicolumn{2}{l}{Total known and unknown dust systematics} & $^{+54.9}_{-45.7}$ & $^{+50.5}_{-14.3}$ & $^{+52.1}_{-14.7}$  \\[1.0 ex]
\hline
\multicolumn{5}{@{}l@{}}{\hbox to 0pt{\parbox{180mm}{\footnotesize
 \par\noindent
 \footnotemark[$a$] ``numerator'' and ``denominator'' refer to those of Equation~4.
    }\hss}}
\end{tabular}\label{sys_tab}
\end{center}
\end{table*}

 \begin{table*}
 \begin{center}
 \caption{The SN~Ia rates in SXDF}
 \begin{tabular}{ccccccccccc}
 \hline
Redshift bin & $z_{eff}$\footnotemark[$a$] & SNR\footnotemark[$b$] & error(stat) & error(sys)\footnotemark[$c$]  & Denom.\footnotemark[$d$] &  N$_{est}$\footnotemark[$e$] & 
N$_{II}$\footnotemark[$f$] & N$_{Ib/c}$\footnotemark[$g$] & N$_{comp}$\footnotemark[$h$] & N$_{Ia}$\footnotemark[$i$] \\[1.0 ex]
\hline
\hline
$0.2 \leq z < 0.6$ & 0.44 & 0.262 & $^{+0.229}_{-0.133}$ & $^{+0.059+0.131}_{-0.120}$ & 13.179 & 4.46 & 0.61 & 1.18 & 0.78 & 3.45$^{+3.02}_{-1.76}$ \\[1.0 ex]
$0.6 \leq z <1.0$ & 0.80 & 0.839 & $^{+0.230}_{-0.185}$ & $^{+0.060+0.419}_{-0.120}$ & 24.199 & 22.48 & 2.52 & 1.88 & 2.23 & 20.31$^{+5.58}_{-4.47}$ \\[1.0 ex]
$1.0 \leq z<1.4$ &  1.14 & 0.705 & $^{+0.239}_{-0.183}$ & $^{+0.102+0.352}_{-0.103}$ & 20.442 & 12.09 & 0.91 & 0.09 & 3.32 & 14.41$^{+4.88}_{-3.75}$ \\[1.0 ex]
 \hline
\multicolumn{10}{@{}l@{}}{\hbox to 0pt{\parbox{180mm}{\footnotesize
 \par\noindent
 \footnotemark[$a$] Effective redshift.
 \par\noindent
 \footnotemark[$b$] Supernova rates are given in units $10^{-4} h_{70}^{3} yr^{-1} Mpc^{-3}.$
 \par\noindent
 \footnotemark[$c$] Know systematic, followed by {\it ad hoc} uncertainty due to dust.
 \par\noindent
 \footnotemark[$d$] Denominator of Equation~4 in units $10^{4} h_{70}^{-3} yr Mpc^{3}.$ 
 \par\noindent
 \footnotemark[$e$] Estimated numbers of observed SNe~Ia (see \S\ref{ratecalc}).
 \par\noindent
 \footnotemark[$f$] Estimated number of contaminating Type~II SNe.
 \par\noindent
 \footnotemark[$g$] Estimated number of contaminating Type~Ib/c SNe after removing two objects classified as SN~Ib/c, i.e., 2-038 and 4-100.
  \par\noindent
 \footnotemark[$h$] Estimated number due to SN typing incompleteness.
  \par\noindent
  \footnotemark[$i$] The number of SN~Ia used in the rate calculation. The errors show statistical uncertainty.
    }\hss}}
  \end{tabular}\label{rate_tab}
\end{center}
\end{table*}

\end{document}